\documentclass[10pt,a4paper]{article}
\usepackage[cp1250]{inputenc}
\usepackage[T1]{fontenc}
\usepackage{amsmath}
\usepackage{amsfonts}
\usepackage{amssymb}
\usepackage{amsthm}
\usepackage{graphicx}
\usepackage{algorithm}
\usepackage{algcompatible}
\usepackage{algpseudocode}
\usepackage{caption}
\usepackage{color}   
\usepackage{hyperref}
\usepackage{breakcites}   

\usepackage{mathptmx} 
\usepackage{enumerate}
\usepackage[misc]{ifsym}
\usepackage{verbatim}
\usepackage{subcaption}
\usepackage{booktabs}
\usepackage{natbib}

\usepackage[left=1in,top=1in,right=1in,bottom=1in,nohead,paperwidth=8.5in, paperheight=11in]{geometry} 

\usepackage{epstopdf}

\theoremstyle{definition}

\title{\bf
	Identifying Important Pairwise Logratios in Compositional Data with Sparse Principal Component Analysis}
	\vspace{3mm}
\author{Viktorie Nesrstov\'a$^{1}$, I. Wilms$^{2}$, K. Hron$^{1}$ and P. Filzmoser$^{5}$}

\date{}  
\begin{document}
\maketitle
\noindent
{\small
$^{1}$Department of Mathematical Analysis and Applications of Mathematics, Palack\'{y} University Olomouc, Faculty of Science, 17. listopadu 12, Olomouc, Czech Republic; \textit{viktorie.nesrstova@gmail.cz} \\
$^{2}$Department of Quantitative Economics, Maastricht University, Tongersestraat 53, Maastricht, The Netherlands\\
$^{3}$Institute of Statistics and Mathematical Methods in Economics, TU Wien, Wiedner Hauptstra\ss e 8-10, Vienna, Austria}

\bigskip

\begin{abstract}
Compositional data  are characterized by the fact that their elemental information is contained in simple pairwise logratios of the parts that constitute the composition. While pairwise logratios are typically easy to interpret, the number of possible pairs to consider quickly becomes (too) large even for medium-sized compositions, which might hinder interpretability in further multivariate analyses. Sparse methods can therefore be useful to identify few, important pairwise logratios (respectively parts contained in them) from the total candidate set. To this end, we propose a procedure based on the construction of all possible pairwise logratios and employ sparse principal component analysis to identify important pairwise logratios. The performance of the procedure is demonstrated both with simulated and real-world data. In our empirical analyses, we propose three visual tools showing (i) the balance between sparsity and explained variability, (ii) stability of the pairwise logratios, and (iii) importance of the original compositional parts to aid practitioners with their model interpretation.
\bigskip


\end{abstract}

\newtheorem{defi}{Definition} [section]   

\newpage

\section{Introduction} \label{Int}

Compositional data (CoDa) are data that carry relative information in their parts (components) \citep{aitchison1986statistical}. 
One can typically think of multivariate observations  measured in units such as percentages or proportions which result in a constant sum (respectively 100 or 1) of their parts. However, the sum of their parts is in fact irrelevant; importantly one should think of CoDa as scale invariant objects \citep{pawlowsky15}. Seminal information in CoDa is contained in the PLRs (PLRs) between the parts of the composition. Even though simple PLRs are  often preferred in multivariate analyses thanks to their straightforward interpretation (as demonstrated in applications; see, e.g., \citealp{greenacre18,greenacre2019variable}), they quickly create a curse of dimensionality. Indeed, considering a composition  of $D$ parts, the number of possible PLRs to consider is $D(D-1)$. This number rapidly increases even for medium-sized compositions, which in turn complicates interpretability in further multivariate analyses. It is therefore desirable to identify few PLRs that are most important (relevant) for the data structure. In this paper, we propose a sparse principal component analysis method to identify such PLRs.

The relative nature of CoDa prevents standard statistical methods from being applied directly. This can, however, be easily overcome by expressing compositions in real logratio coordinates. A popular choice in this context are orthonormal logratio coordinates (olr), with the special case of balance coordinates being  promoted in the last years \citep{pawlowsky15}. These coordinates are constructed such that they express the dominance of one group of compositional parts with respect to another group. They are built based on the procedure called sequential binary partition (SBP) where non-overlapping groupings of parts are gradually derived. However, to build a meaningful SBP, prior expert knowledge of the data is usually required. But such knowledge could be particularly complicated to acquire for high-dimensional compositions as the dimensionality 
affects both the construction of the SBP and interpretation of balance coordinates.
While there exist data-driven algorithms to overcome this obvious handicap \citep{martin2018advances}, an appealing alternative is to analyze CoDa through simple PLRs that typically greatly facilitate interpretability.

Since the dimensionality of a $D$-part composition is only $D-1$ \citep{pawlowsky15}, analyzing all PLRs simultaneously results into a positive semidefinite covariance matrix which causes problems and possible inconsistencies in many statistical methods. Fortunately this is not the case (amongst others) for Principal Component Analysis (PCA) where using either all PLRs or any olr coordinates (or centred logratio coefficients, probably the most popular in this context, see \citealp{AG02}) lead to the same score values, and therefore, also the same loadings representing PLRs \citep{daunis11, hron2021analysing}. Directly using all PLRs as a proper representation of compositions is also preferred when affine equivariance (or even just orthogonal equivariance as for PCA) is lacking, see for example \cite{stefelova21} in the context of cellwise outlier detection or \cite{tolosana19} in the context of machine learning. Nevertheless, not all PLRs are necessarily relevant for the multivariate data structure. Important information on the underlying processes in the data could be contained in few distinct pairs, consisting of certain key parts. It would therefore be desirable to primarily identify  relevant logratios which can then be used for further analysis, or even for the construction of interpretable logratio coordinates.

There have been several attempts to design a method that carefully selects PLRs, see for instance \cite{greenacre18, greenacre2019variable}, whose procedure is based on the construction of relevant, interpretable PLRs that build up a logratio coordinate system.
Our aim in this paper is not to build the whole coordinate system out of the selected PLRs, but instead to focus on those PLRs that capture most information in the data. 
We therefore develop a new procedure in the context of PCA for compositional data by using sparse PCA to identify the most relevant logratios (parts). 
We propose to use the sparse PCA procedure introduced in \cite{erichson2020sparse}, namely sparse PCA via variable projection, as it provides an effective and stable algorithm which is able to deal with the high-dimensionality of PLRs. 

The manuscript is structured as follows: 
In Sect. \ref{method}, we revise PCA for CoDa expressed in PLRs and introduce the sparse PCA method 
to identify important PLRs.
A simulation study is provided in Sect. \ref{Sim} and two geochemical data sets are analyzed in Sect. \ref{real}. 
Here, we also present various graphical tools to aid practitioners with their model interpretation.
The final Sect. \ref{End} ends with some concluding remarks.

\section{Sparse PCA with pairwise logratios} \label{method}
We start by revising Principal Component Analysis (PCA) for CoDa expressed in PLRs in Sect. \ref{subsec:pca}. In Sect. \ref{subsec:sparsepca} we explain how one can use sparse PCA to identify important PLRs from the total candidate set.

\subsection{Principal Component Analysis with CoDa} \label{subsec:pca}
Principal Component Analysis (PCA) is a well-known dimension reduction method for multivariate data, which is also commonly used as exploratory data analysis tool with CoDa. The key idea of PCA is to reduce the dimensionality of the original data set by constructing a
new set of latent variables, called Principal Components (PCs),
that are linear combinations of the original variables and  mutually uncorrelated. 
PCs are constructed such that the first PC maximizes the variance in the original data and subsequent PCs maximize the remaining variance. 
For CoDa, all computations need to be done in orthonormal coordinates, and we refer to \cite{filzmoser2018applied} for further details on PCA with CoDa.

Let  $\mathbf{X}=(x_{ij})_{1\leq i \leq n, 1\leq j \leq D}$ be a matrix carrying information about a $D$-part composition (in the columns) with $n$ observations (in the rows). We represent the CoDa through a matrix of
PLRs with $\frac{D(D-1)}{2}$ columns (as the information carried in $\textrm{ln}\frac{x_{i}}{x_{j}}$ is the same as in $\textrm{ln}\frac{x_{j}}{x_{i}}$), which will be further noted as $\mathbf{X}_{pair}$.
It is assumed that the matrix $\mathbf{X}_{pair}$ is centered prior to performing PCA.

To construct the PCs, we adopt the notation of \cite{erichson2020sparse}, to be further used in Sect. \ref{subsec:sparsepca}. The $i$th ($i=1,\ldots, \frac{D(D-1)}{2}$) principal component score $\mathbf{z}_{i}$ is the $n$-dimensional (column) vector formed as a linear weighted combination of all PLRs
\begin{equation}
    \mathbf{z}_{i} = \mathbf{X}_{pair} \mathbf{p}_{i},
\end{equation}
where $\mathbf{p}_{i}$ is a $\frac{D(D-1)}{2}$-dimensional (column) vector of loadings.  
All PCs $\mathbf{z}_{1}, \ldots,  \mathbf{z}_{\frac{D(D-1)}{2}}$ can then be combined in an $n\times \frac{D(D-1)}{2}$ score matrix $\mathbf{Z}$ given by
\begin{equation}
    \mathbf{Z} = \mathbf{X}_{pair} \mathbf{P}.
\end{equation}
Note that, apart from the fact that only up to $D-1$ PCs have nonzero variance, one typically does not consider all but instead only the first few to capture a desirable fraction of the total variability in the original data. 

After having obtained the PCA model, one can proceed to visualization in a compositional biplot, i.e.\ a graphical display of both variables (loadings) and observations (scores). As such, one can  assess the importance of each PLR, group of logratios, or observation for explaining the total variability, see for instance \cite{daunis11}. 

However, as the dimensionality of the composition increases, it becomes cumbersome to identify those logratios or even key original components that are most relevant for revealing leading processes in the data. 
This calls for the need of a regularization technique that can identify such important logratios. To this end, we resort to sparse PCA, as discussed in the next section.

\subsection{Selecting Important pairwise logratios via Sparse PCA} \label{subsec:sparsepca}
We use the sparse PCA method, based on variable projection, of \cite{erichson2020sparse} to encourage sparsity in the PCA loadings of all $\frac{D(D-1)}{2}$ PLRs.
This method is directly based on the input data matrix (instead of a covariance matrix), and it can cope with high-dimensional data having more variables than observations.
The method aims to find a set of sparse loading vectors, i.e.\ vectors with just a few non-zero elements, by minimizing the objective function 
\begin{equation} \label{obj_sparsepca}
\min_{\mathbf{H},\mathbf{B}} \ \frac{1}{2} \lVert \mathbf{X} - \mathbf{X}\mathbf{B}\mathbf{H}^T \rVert_{F}^{2} + \psi(\mathbf{B}),\quad \textrm{subject to}\quad \mathbf{H}^T\mathbf{H} = \mathbf{I},    
\end{equation}
where $\mathbf{X}$ is the data matrix (to be clarified below for the CoDa case),
$\mathbf{B}$ is a sparse loadings (weight) matrix, $\mathbf{H}$  an orthonormal matrix and $\psi$ denotes a sparsity inducing penalty function. For the latter, we take an elastic net penalty 
as given by 
\begin{equation}
\psi(\mathbf{B}) = \alpha\lVert\mathbf{B}\rVert_{1} + \frac{1}{2}\beta\lVert \mathbf{B} \rVert_{2} ^2,\quad
\end{equation}
where $\alpha, \beta >0$ are tuning parameters. The tuning parameter $\alpha$ controls the degree sparsity of the resulting loading vectors: the higher its value, the sparser the result. For the tuning parameter $\beta$, we take the default setting of $\beta=0.0001$ as used in \cite{erichson2020sparse} \footnote{One can use a simple lasso penalty instead of an elastic net penalty by setting $\beta = 0$. We verified that similar results, compared to the ones reported in the paper, are obtained in such a case.}.
The resulting estimated PCs are then defined as $\mathbf{Z} = \mathbf{X}\widehat{\mathbf{B}}$.   
While standard PCA is orthogonal equivariant, meaning that any olr coordinates or set of all $\frac{D(D-1)}{2}$ PLRs would lead to the same scores, this is no longer the case with sparse PCA because of the $\ell_1$-regularization. Using all $\frac{D(D-1)}{2}$ PLRs as seminal information in CoDa for PCA is therefore not only a natural choice with respect to the aim of our analysis, but also an essential requirement to obtain consistent results.

The  procedure to perform sparse PCA for CoDa expressed in PLRs can then be summarized into the following two simple steps:
\begin{enumerate}
    \item Construct a matrix of all
    $\frac{D(D-1)}{2}$ PLRs, denoted as $\mathbf{X}_{pair}$.
    \item  Perform sparse PCA  by minimizing the objective function \eqref{obj_sparsepca} with the matrix $\mathbf{X}_{pair}$, and let $\widehat{\mathbf{B}}$ denote the estimated loadings matrix.
\end{enumerate}

The sparse PCA procedure thus results in a sparse loading matrix $\widehat{\mathbf{B}}$ of dimension 
$\frac{D(D-1)}{2} \times \frac{D(D-1)}{2}$.
The importance of each PLR can then be studied from the sparsity pattern in the corresponding row of the loadings matrix. 

In the remainder, we focus on the first two PCs. Typically, these are used in the construction of the standard biplot, and also retain the most important variability. 
The loadings matrix then has only two columns to be inspected.
In the remainder, we talk about logratios with zero loadings (in short ``zero logratios") as those logratios having zero loadings in both the first and second PC. 

All calculations were performed in R \citep{Rsoft}. To perform sparse PCA as described above, the function \texttt{spca} available in package \texttt{sparsepca} \citep{spcaLib} was used. The code of our algorithm is available on the GitHub page of the first author (\url{https://github.com/NesrstovaV/PairwiseLogrs-sPCA.git}).

\section{Simulation Study} \label{Sim}
We perform a simulation study to investigate the ability of the proposed method to identify important PLRs among the total set.
In Sect. \ref{subsec:sim-scenarios}, we describe the simulation scenarios. In Sect. \ref{subsec:sim-spca-vs-step} we compare our sparse PCA proposal against the STEP method of \cite{greenacre2019variable} before further zooming into the performance of sparse PCA in Sect. \ref{subsec:sim-spca}.

\subsection{Simulation Scenarios} \label{subsec:sim-scenarios}
We consider a data generating process via balance coordinates consisting of a combination of ``relevant" and ``noise" balances. 
As such, we embed ``sparsity" in the compositional parts since compositional parts in the relevant coordinates are considered as important, whereas parts in the noise coordinates are unimportant and hence should display zero loadings in the sparse PCA procedure.

We consider three simulation scenarios, each of them in combination with either $D=10$ parts or $D=20$ parts. Table \ref{Table ABC} summarizes the three simulation scenarios with an SBP scheme for a $D=10$-part composition with 9 balances. 
An SBP is usually encoded
in a sign matrix, which illustrates the division of parts into groups as can be seen from Table
\ref{Table ABC}. Typically, parts in the numerator are coded as ``+" and parts in the denominator as ``-".
Parts that are not included in a particular balance are coded as ``0".
The scenarios in Table
\ref{Table ABC} differ in terms of the proportion of relevant versus noise balances.
In scenario A, there is a (roughly) equal amount of relevant ($b_1-b_5$) and noise ($b_6-b_9$) balances in the coordinate representation.
In contrast, in scenario B the relevant balances ($b_1-b_7)$ dominate, while in scenario C the noise balances ($b_3-b_9)$ dominate.

\begin{table}
\caption{SBP tables of the different scenarios for $D=10$.
Cells with ``$+$'' (``$-$'') sign mark parts in the numerator (denominator) of the balance coordinate.
Zeros mark parts not included in the balance.
}
\begin{minipage}{\linewidth}
\centering
{\small{\tabcolsep 6pt\renewcommand{\arraystretch}{0.8}\small
\begin{tabular}{|c|cccccccccc|} \hline
 & $x_{1}$ & $x_{2}$ & $x_{3}$ & $x_{4}$ & $x_{5}$ & $x_{6}$ & $x_{7}$ & $x_{8}$ & $x_{9}$ & $x_{10}$ \\ \hline
$b_{1}$ & + & + & + & + & + & - & - & - & - & - \\
$b_{2}$ & + & + & + & + & - & 0 & 0 & 0 & 0 & 0 \\
$b_{3}$ & + & + & + & - & 0 & 0 & 0 & 0 & 0 & 0 \\
$b_{4}$ & + & + & - & 0 & 0 & 0 & 0 & 0 & 0 & 0 \\
$b_{5}$ & + & - & 0 & 0 & 0 & 0 & 0 & 0 & 0 & 0 \\ \hline
$b_{6}$ & 0 & 0 & 0 & 0 & 0 & + & - & - & - & - \\
$b_{7}$ & 0 & 0 & 0 & 0 & 0 & 0 & + & - & - & - \\
$b_{8}$ & 0 & 0 & 0 & 0 & 0 & 0 & 0 & + & - & - \\
$b_{9}$ & 0 & 0 & 0 & 0 & 0 & 0 & 0 & 0 & + & - \\
\hline
\end{tabular}}}
\subcaption{Scenario A: relevant balances $b_{1}$--$b_{5}$, noise balances $b_{6}$--$b_{9}$.} \label{table A}
\end{minipage} 
\begin{minipage}{\linewidth}
\centering
{\small{\tabcolsep 6pt\renewcommand{\arraystretch}{0.8}\small
\begin{tabular}{|c|cccccccccc|} \hline
 & $x_{1}$ & $x_{2}$ & $x_{3}$ & $x_{4}$ & $x_{5}$ & $x_{6}$ & $x_{7}$ & $x_{8}$ & $x_{9}$ & $x_{10}$ \\ \hline
$b_{1}$ & + & + & + & + & + & + & + & - & - & - \\
$b_{2}$ & + & + & + & + & + & + & - & 0 & 0 & 0 \\
$b_{3}$ & + & + & + & + & + & - & 0 & 0 & 0 & 0 \\
$b_{4}$ & + & + & + & + & - & 0 & 0 & 0 & 0 & 0 \\
$b_{5}$ & + & + & + & - & 0 & 0 & 0 & 0 & 0 & 0 \\ 
$b_{6}$ & + & + & - & 0 & 0 & 0 & 0 & 0 & 0 & 0 \\ 
$b_{7}$ & + & - & 0 & 0 & 0 & 0 & 0 & 0 & 0 & 0 \\ \hline
$b_{8}$ & 0 & 0 & 0 & 0 & 0 & 0 & 0 & + & - & - \\
$b_{9}$ & 0 & 0 & 0 & 0 & 0 & 0 & 0 & 0 & + & - \\
\hline
\end{tabular}}}
\subcaption{Scenario B: relevant balances $b_{1}$--$b_{7}$, noise balances $b_{8}$ and $b_{9}$.} \label{table B}
\end{minipage}%
\\
\begin{minipage}{\linewidth}
\centering
{\small{\tabcolsep 6pt\renewcommand{\arraystretch}{0.8}\small
\begin{tabular}{|c|cccccccccc|} \hline
 & $x_{1}$ & $x_{2}$ & $x_{3}$ & $x_{4}$ & $x_{5}$ & $x_{6}$ & $x_{7}$ & $x_{8}$ & $x_{9}$ & $x_{10}$ \\ \hline
$b_{1}$ & + & + & - & - & - & - & - & - & - & - \\
$b_{2}$ & + & - & 0 & 0 & 0 & 0 & 0 & 0 & 0 & 0 \\ \hline
$b_{3}$ & 0 & 0 & + & - & - & - & - & - & - & - \\  
$b_{4}$ & 0 & 0 & 0 & + & - & - & - & - & - & - \\
$b_{5}$ & 0 & 0 & 0 & 0 & + & - & - & - & - & - \\ 
$b_{6}$ & 0 & 0 & 0 & 0 & 0 & + & - & - & - & - \\ 
$b_{7}$ & 0 & 0 & 0 & 0 & 0 & 0 & + & - & - & - \\
$b_{8}$ & 0 & 0 & 0 & 0 & 0 & 0 & 0 & + & - & - \\
$b_{9}$ & 0 & 0 & 0 & 0 & 0 & 0 & 0 & 0 & + & - \\
\hline
\end{tabular}}}
\subcaption{Scenario C: relevant balances $b_{1}$ and $b_{2}$, noise balances $b_{3}$--$b_{9}$.} \label{table C}
\end{minipage}   
\label{Table ABC}
\end{table}

In addition to a $D=10$-part composition, we also consider $D=20$ with 10 (scenario A), 15 (scenario B) and 4 (scenario C) relevant balances; the SBP was performed analogously to the case for $D=10$. While these compositions appear to be of moderate size at first sight, the number of PLRs in each is considerable, namely  90 and 380 respectively (or resp. 45 and 190 when logratios are considered up to sign).

The relevant and noise balance coordinates are generated such that their variability reflects settings that typically occur in CoDa data sets.
Specifically, the relevant balances are drawn from a multivariate normal distribution
with mean zero and covariance matrix with ones on the diagonal and 0.7 for the off-diagonal elements.
The noise balances are drawn from a uniform distribution $U(-2,2)$.
We then back-transform the system of all balance coordinates to obtain an  $n\times D$  data matrix $\mathbf{X}$  of raw compositions.
We fix the number of observations to $n=100$ and  use 100 simulation runs.

\subsection{Comparing Sparse PCA to STEP} \label{subsec:sim-spca-vs-step}
We start by comparing the proposed sparse PCA approach to the method of~\cite{greenacre2019variable}. The latter is called stepwise ratio selection (STEP) and is available in the function \texttt{STEP} of the R package \texttt{easyCODA} \citep{easyCODA}. 
STEP aims to obtain a set of PLRs that serves as a suitable representation of the whole compositional data set. 
As STEP constructs logratio coordinates, it chooses at maximum $D-1$ PLRs (i.e.\ coordinates). 
Therefore, we compare our proposal to STEP through $D-1$ PLRs, the maximum we can obtain with the STEP algorithm.
STEP orders the logratios according to their percentage of explained variance, with the first one having the highest percentage of explained variability.
Moreover, the logratios are chosen such that they are not linearly dependent on each other. For example, if logratios $\textrm{ln}\frac{x_{i}}{x_{j}}$ and $\textrm{ln}\frac{x_{j}}{x_{k}}$ are chosen, then the logratio $\textrm{ln}\frac{x_{i}}{x_{k}}$ can not longer be selected.

\begin{figure}[t]
\includegraphics[width=1\textwidth]{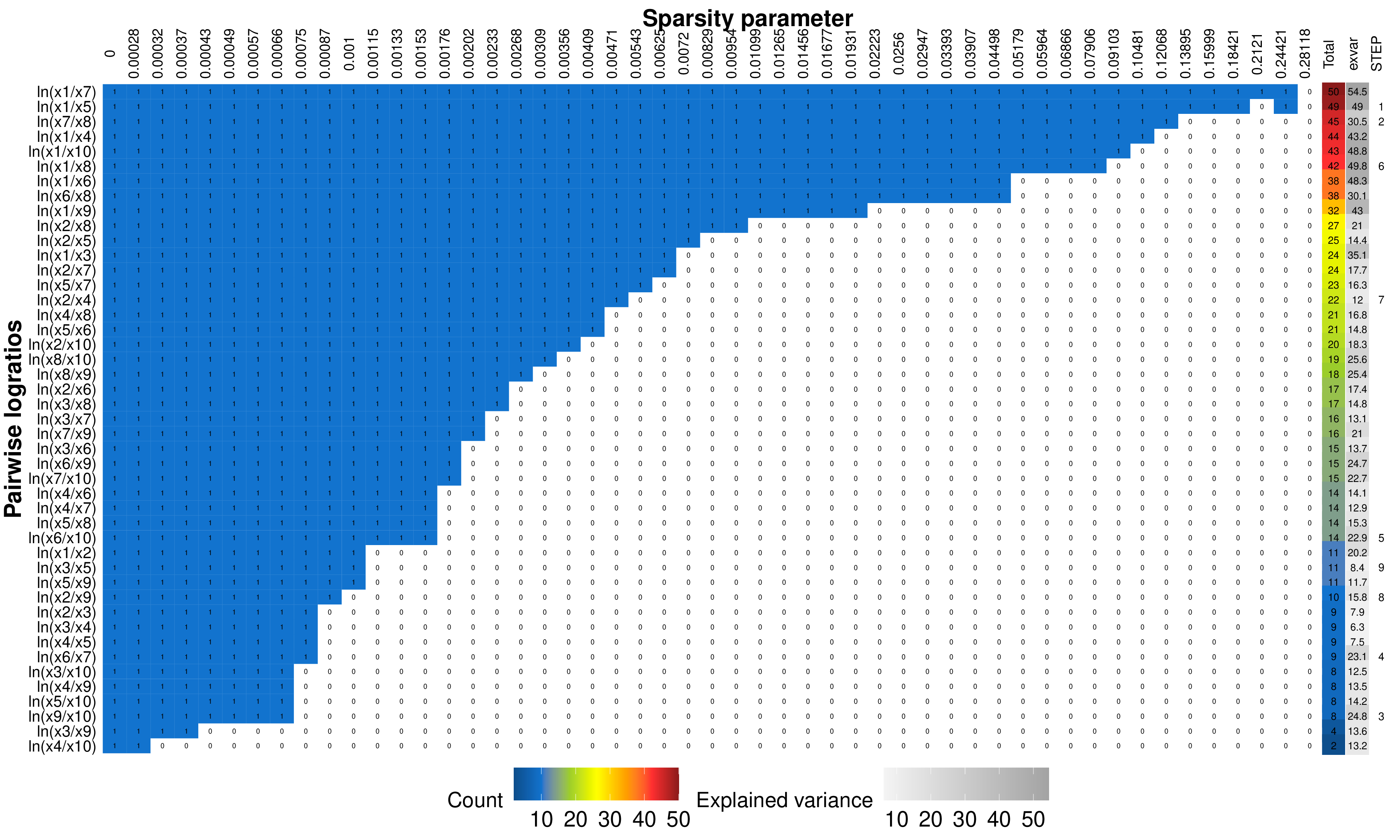} 
  \caption{Stability plot for an illustrative example of simulation scenario A with $D=10$. The graph shows stability paths for PLRs (in rows) where colored cells with ones reflect selection of the PLR (i.e.\ non-zero loading) for the sparse PCA solution with particular $\alpha$ (in columns, until full sparsity is reached). The column ``total" displays the number of models (row counts) out of 51 in which the PLR is selected, the column ``exvar" gives the explained variability by the logratio. The last column ``STEP" shows the ranks of $D-1$ PLRs selected by STEP.}
  \label{fig:Sim1Aheat}
\end{figure}

To select $D-1$ logratios for our proposal, we first sort (in each simulation run) the PLRs according to their stability, and then keep only the first $D-1$ PLRs for comparison with STEP. 
A logratio is considered to be more stable than another if it remains selected for higher values of the sparsity parameter $\alpha$.  
We apply the sparse PCA method for different values of $\alpha$, ranging from the solution with no sparsity ($\alpha=0$) towards maximal sparsity ($\alpha_{\text{max}}$, which varies from setting to setting). We consider a logarithmic-spaced grid of $51$ sparsity parameters between these extreme values and apply the sparse PCA procedure for each value in the grid.
We then display stability paths (see, e.g., Figure \ref{fig:Sim1Aheat}), similar to regularization paths in the context of sparse regression analysis.

Figure \ref{fig:Sim1Aheat} gives an illustrative example (namely for one simulation run for scenario~A, $D = 10$) of the stability of each logratio for increasing values of the sparsity parameter $\alpha$ on the horizontal axis. On the vertical axis we show the $\frac{D(D-1)}{2}$ PLRs.\footnote{We plot only $\frac{D(D-1)}{2}$  PLRs to avoid redundancy. For example, logratios $\textrm{ln}\frac{x_{1}}{x_{5}}$ and $\textrm{ln}\frac{x_{5}}{x_{1}}$ differ just by sign, and the percentage of variability explained by each is therefore the same.} 
The cells then highlight the selection results: ones in  blue cells correspond to selected logratios (i.e.\ with non-zero loading); zeros in white cells correspond to not-selected logratios (i.e.\ with zero loading). 
We order the PLRs according to the number of times they are selected across all the grid points of the sparsity parameter (see column ``Total" in Fig. \ref{fig:Sim1Aheat}). 
The most stable logratios thus appear at the top and remain non-zero even in very sparse models, as can be seen from the longer blue stretches.
The column ``exvar" shows explained variability by each PLR. 
Finally, the last column ``STEP" shows the ranks of the $D-1$ PLRs as selected by the STEP algorithm. 

Once we have a set of $D-1$ selected PLRs for our proposal as well as for STEP, we can then investigate their performance. We compare our sparse PCA proposal to STEP on two criteria, namely
(i) the percentage of important PLRs chosen by each method, and  (ii) the rank of the chosen logratios based on their variances.

\begin{figure}
\centering

\begin{subfigure}{0.475\textwidth}    
\centering
\includegraphics[width=\textwidth]{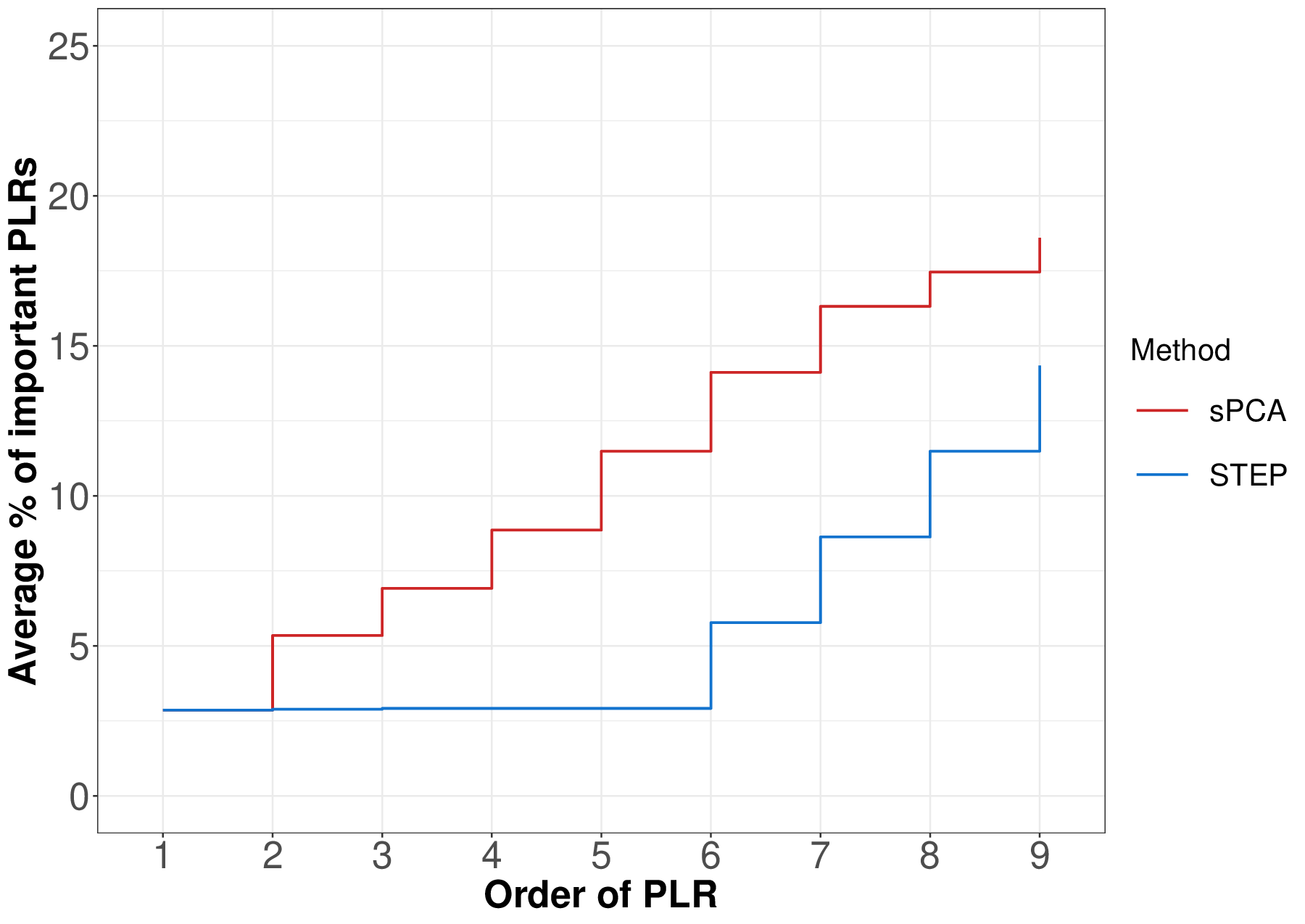}
\caption{Scenario A}
\label{1subfig:2b}
\end{subfigure}
\begin{subfigure}{0.475\textwidth}   
\centering
\includegraphics[width=\textwidth]{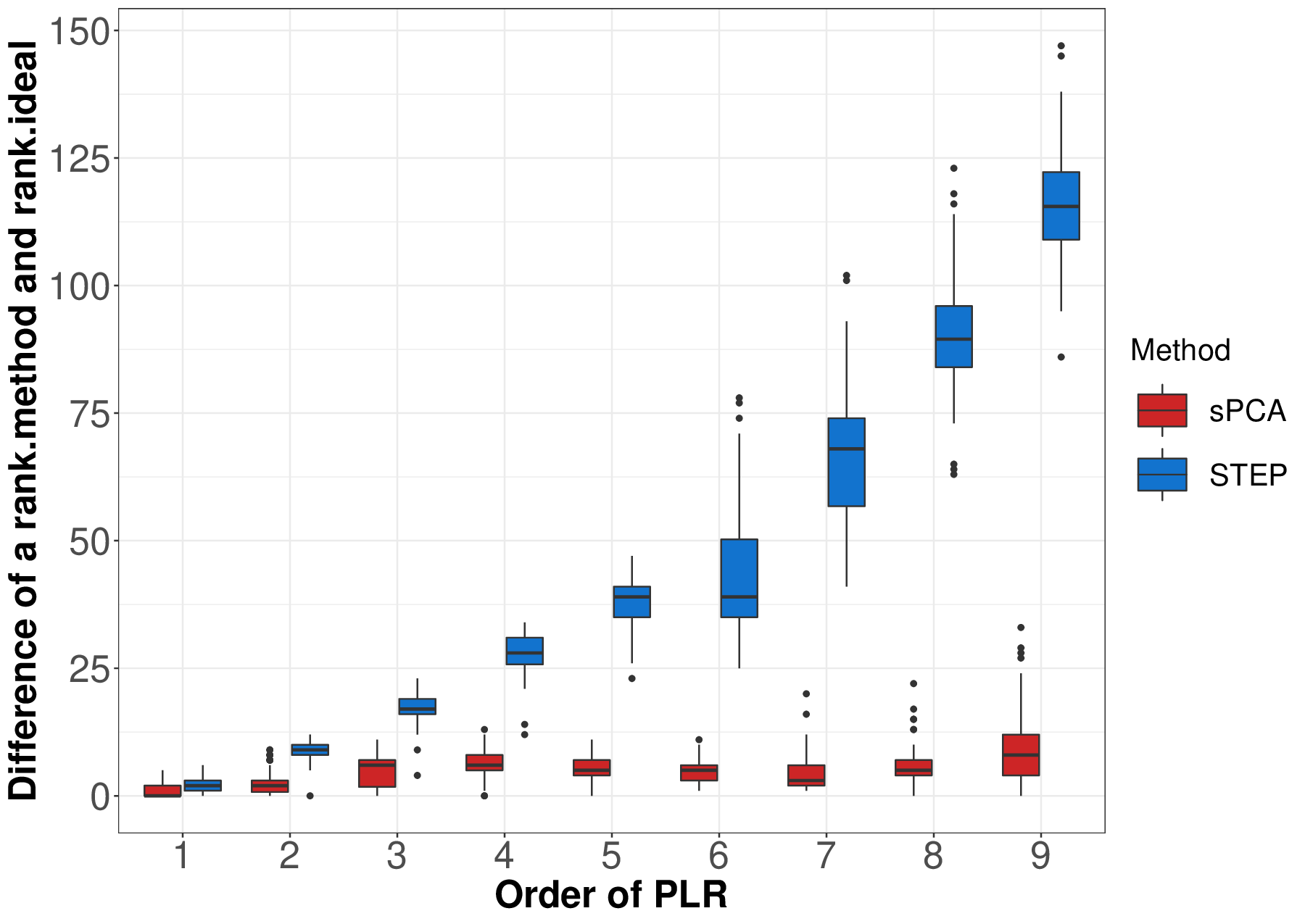}
\caption{Scenario A}
\label{1subfig:2a}
\end{subfigure}%

\begin{subfigure}{0.475\textwidth}    
\centering
\includegraphics[width=\textwidth]{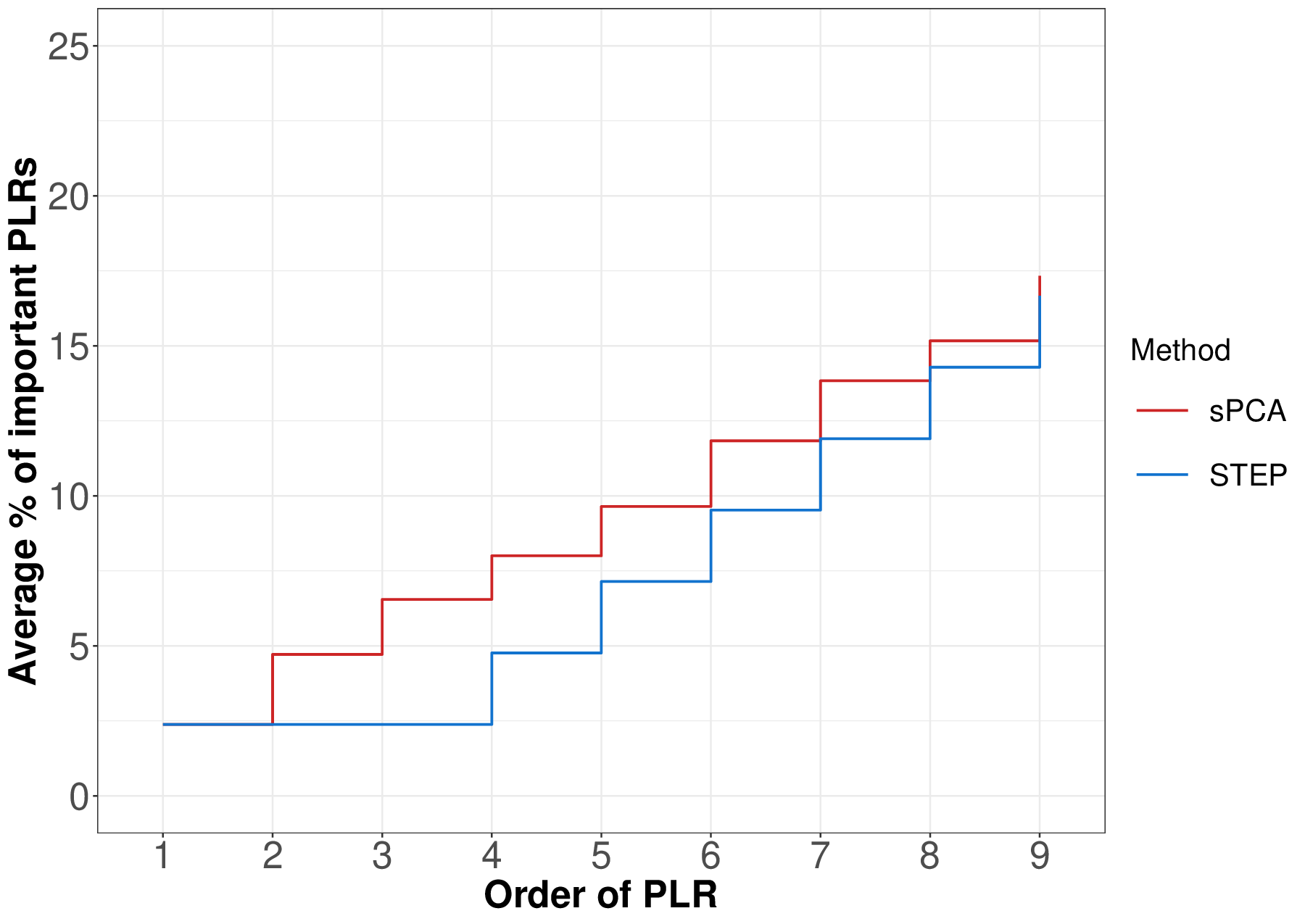}
\caption{Scenario B}
\label{1subfig:2d}
\end{subfigure}
\begin{subfigure}{0.475\textwidth}    
\centering
\includegraphics[width=\textwidth]{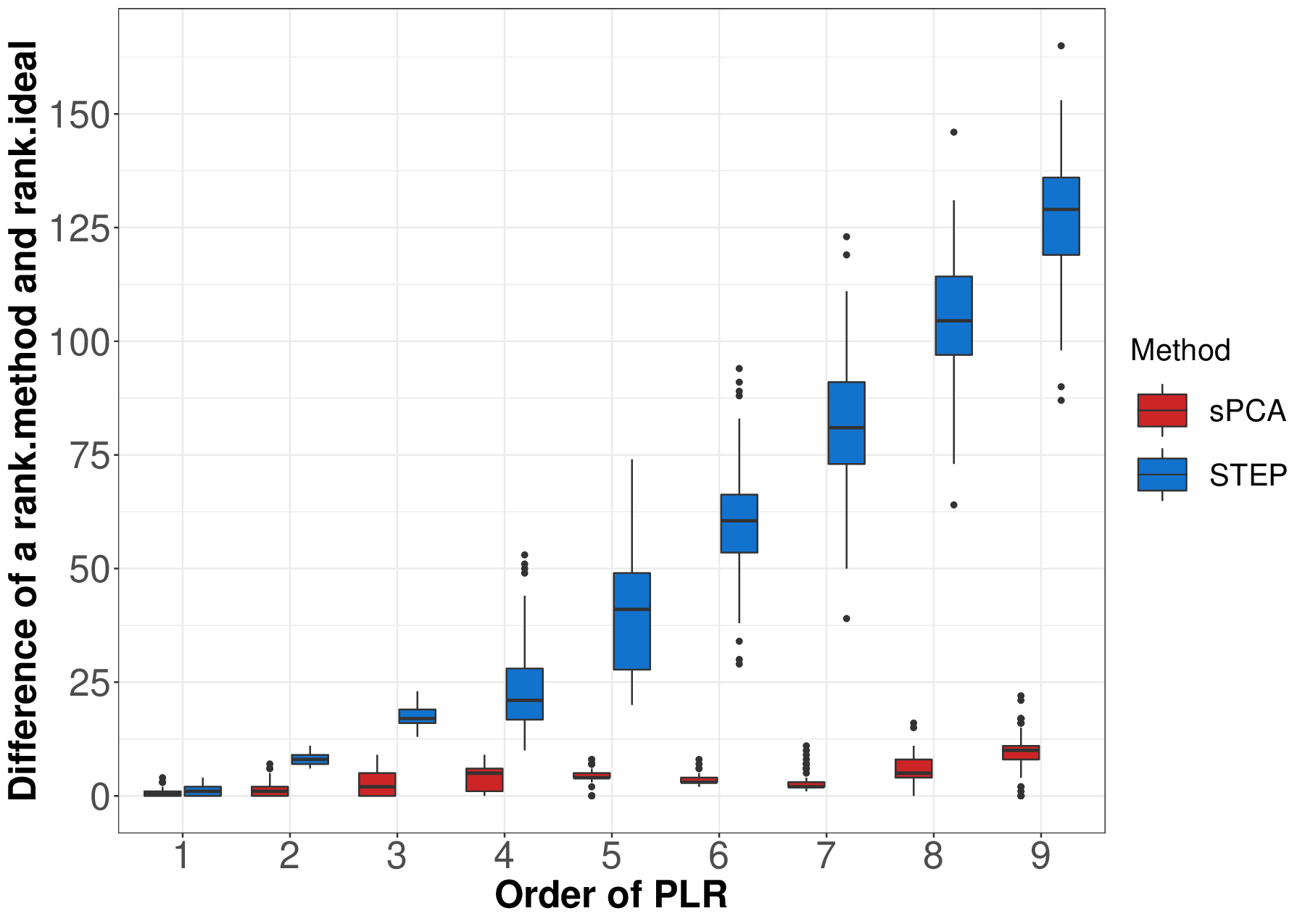}
\caption{Scenario B}
\label{1subfig:2c}
\end{subfigure}

\begin{subfigure}{0.475\textwidth}   
\centering
\includegraphics[width=\textwidth]{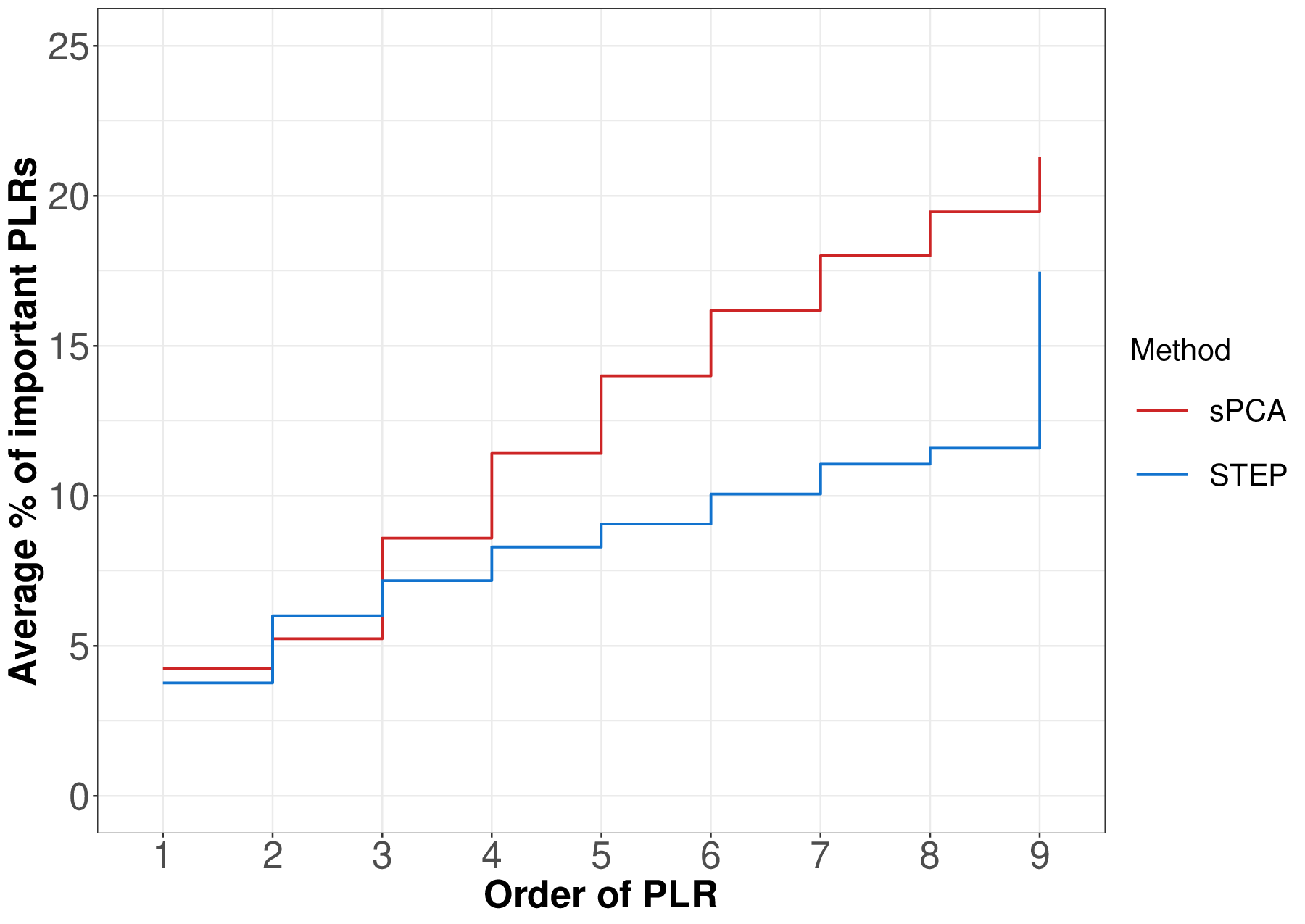}
\caption{Scenario C}
\label{1subfig:2f}
\end{subfigure}%
\begin{subfigure}{0.475\textwidth}   
\centering
\includegraphics[width=\textwidth]{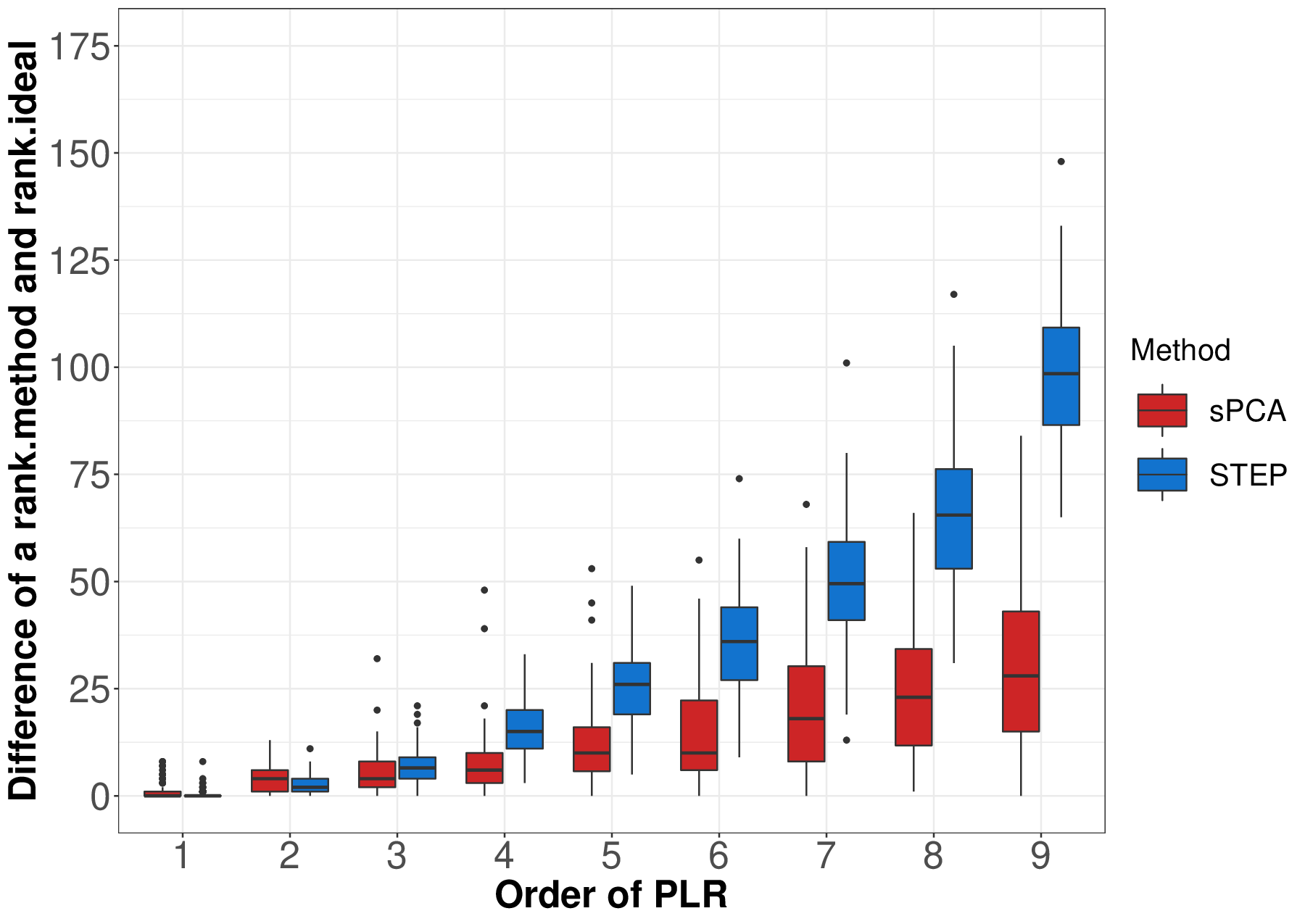}
\caption{Scenario C}
\label{1subfig:2e}
\end{subfigure}%

\caption{Comparison of sparse PCA to STEP on simulation scenarios A, B and C for $D=10$. 
Left: average percentage of correctly identified important PLRs for sPCA (red) and STEP (blue) for the first $D-1$ ordered PLRs (PLR) on the horizontal axis.
Right: boxplots showing the differences between the cumulative ranks of the PLR obtained from sPCA (red) or STEP (blue) and the  ideal ranks. }
\label{fig:LogrsP}
\end{figure}

{\bf Selection of important logratios.}
Given the set of $D-1$ PLRs obtained from sparse PCA (i.e.\ the most stable ones) and STEP, we can investigate whether these belong to the set of important PLRs (corresponding to the relevant balances).
In the left panels of Fig. \ref{fig:LogrsP}, we provide step plots that display the average percentage, across simulation runs, of important PLRs that are picked up in the first $D-1$ selected PLRs of sparse PCA (sPCA, in red) and STEP (in blue).
We display the results on the different simulation scenarios for $D=10$ in respective panels of Fig. \ref{1subfig:2b}, \ref{1subfig:2d} and\ref{1subfig:2f}.
Results for all scenarios of $D = 20$ are reported and summarized in Appendix \ref{appA}, as these are very similar.

Sparse PCA is, overall, more successful in capturing the important PLRs than STEP, since the increase in percentage of correctly captured important logratios occurs faster than for STEP.
The discrepancy between both is the largest for scenario A. In scenario C, where the data contain few important PLRs, both methods perform similarly for the first two chosen logratios, but further on sPCA outperforms STEP again. 
Importantly, sparse PCA, overall, does not perform worse than STEP. 
It is therefore a suitable alternative for choosing important PLRs, if a complete coordinate system formed by PLRs is not a primary goal, even more so since STEP is limited by the number of chosen logratios ($D-1$) whereas our proposal is not.

{\bf  Ranking based on variances.}
Next, we compare sparse PCA and STEP based on the ranks of the selected PLRs. 
We start by computing the ``ideal rank" of each 
PLR, thereby ranking them according to their variances with the logratio having the highest variance being ranked first.
For sparse PCA, we obtain the first $D-1$ logratios ordered according to stability, for STEP we simply obtain the $D-1$ ordered logratios.
For each method, we then compute the cumulative ranks for the selected logratios, going from 1 up until $D-1$. In case of sPCA, we first give rank to all $\frac{D(D-1)}{2}$ logratios (ordered by stability) and then keep just the first $D-1$ most stable ones. 
Finally, we obtain the difference between these cumulative sum of ranks and the ideal cumulative sum of rank. 
By definition, this difference can only be positive.
If the difference is zero, the logratios ordered by the method exactly correspond to logratios ranked by their variance. 
The larger the difference, the higher the tendency of the method to rank logratios with lower variances before logratios with higher ones.

Boxplots in the right panel of Fig. \ref{fig:LogrsP} show the variability of these differences across the simulation runs, and this for the different simulation scenarios.
Sparse PCA considerably outperforms STEP as the difference with the ideal ranks is smaller and the margin by which sparse PCA outperforms STEP gradually increases as more PLRs are considered.
This implies that logratios with high variances are typically appearing as the most stable ones in our sparse PCA procedure, which is a desirable feature.

Given the good performance of our sparse PCA proposal against STEP, we further zoom into its performance across the different simulation scenarios in the next section.

\subsection{Performance of Sparse PCA across the Simulation Scenarios} \label{subsec:sim-spca}

We discuss the performance of sparse PCA in terms of (i) trading-off explained variability with model parsomony and (ii) ability to correctly separate the important logratios (corresponding to the relevant balances) from the unimportant logratios (corresponding to the noise balances). To this end, we compute the 
false positive rate (FPR) and false negative rate (FNR) \citep{fawcett2006introduction} given by
$$
FPR  = \frac{FP}{FP + TN} \qquad FNR = \frac{FN}{FN+TP}
$$
where the true positive $TP$ (true negative $TN$) indicate the number of true important, non-zero (unimportant, zero) logratios. The false positives $FP$ (false negatives $FN$) then give the number of truly unimportant, zero (important non-zero) logratios that are incorrectly given a non-zero (zero) loading.

We are now ready to discuss the results for each of the three simulation scenarios with $D=10$. 
For $D=20$, the main findings remain unchanged; we therefore report and summarize these results in Appendix \ref{appA}.

\begin{figure}
\centering
\begin{subfigure}{0.475\textwidth}   
\centering
\includegraphics[width=\textwidth]{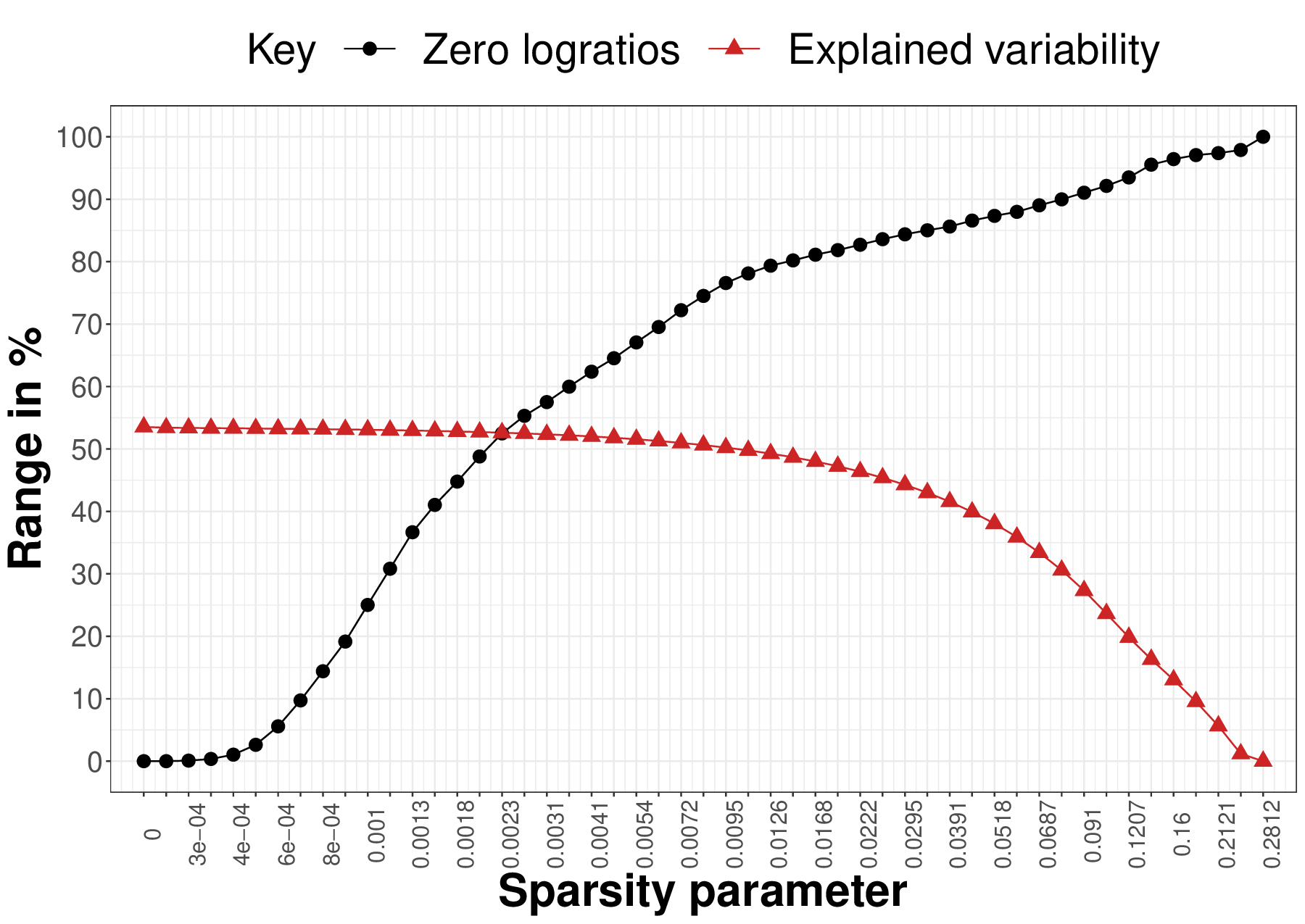}
\caption{Development of sparsity with increasing $\alpha$}
\label{subfig:1a}
\end{subfigure}%
\begin{subfigure}{0.475\textwidth}    
\centering
\includegraphics[width=\textwidth]{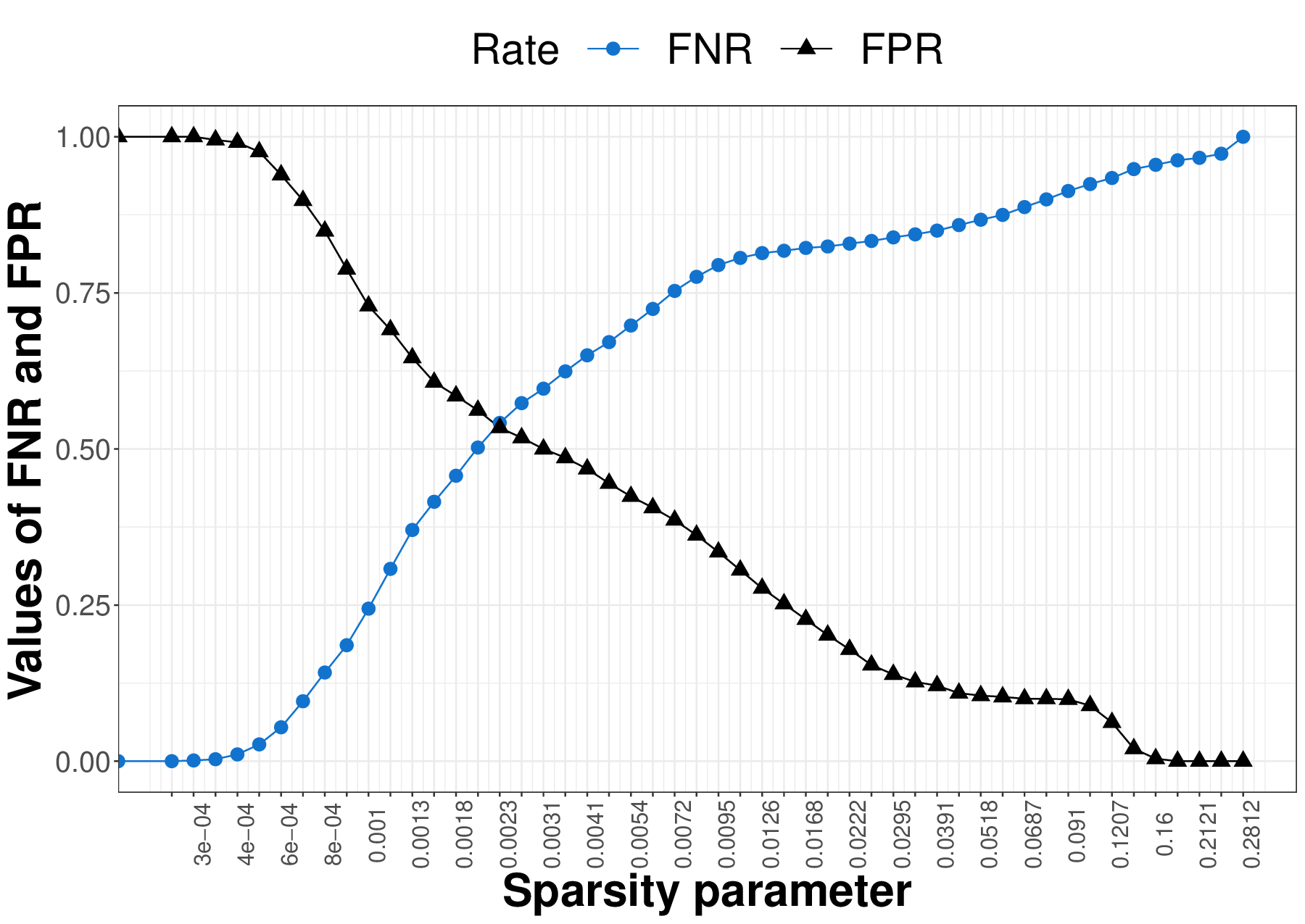}
\caption{FNR and FPR with increasing $\alpha$}
\label{subfig:1b}
\end{subfigure}
\caption{Simulation scenario A for $D=10$. Left: percentage of zero logratios (black)  and explained variability (red) for different values of the sparsity parameter $\alpha$. Right: FPR (black) and FNR (blue)  for different values of $\alpha$. }
\label{fig:SimA}
\end{figure}

{\bf Results Scenario A.}
Figure \ref{fig:SimA} shows the results for simulation scenario A with $D = 10$ parts. 
Figure \ref{subfig:1a} displays the percentage of explained variability (in red)  and percentage of zero logratios (in black) across the different values of the sparsity parameter $\alpha$. Note that for $\alpha=0$, the percentage of explained variability does not start at 100\% since we only focus on the first two PCs.
When gradually increasing the sparsity parameter, we see that the percentage of explained variability only mildly decreases whereas the percentage of zero logratios increases much quicker. This trade-off provides interesting opportunities to consider sparser solutions with  (few) important logratios. Indeed, the dense, hard-to-interpret solution ($\alpha=0$) offers the highest attainable percentage of explained variability (53.5\%), but a considerably sparser solution with up to 52.6\% 
zero logratios (around $\alpha = 0.0023$) only pays a price in terms of explained variability of around 0.9\% (i.e.\ drop from 53.5\% to 52.6\% in explained variability).
The PCA solution with that same sparsity parameter attains the best possible balance between the FNR and FPR which are both roughly at 0.5, as can be seen from Fig. \ref{subfig:1b}. Note that, by definition, the FPR decreases from 1 to 0 as the degree of sparsity increases from zero to maximal sparsity, while the opposite occurs for the FNR.

\begin{figure}
\centering
\begin{subfigure}{0.475\textwidth}   
\centering
\includegraphics[width=\textwidth]{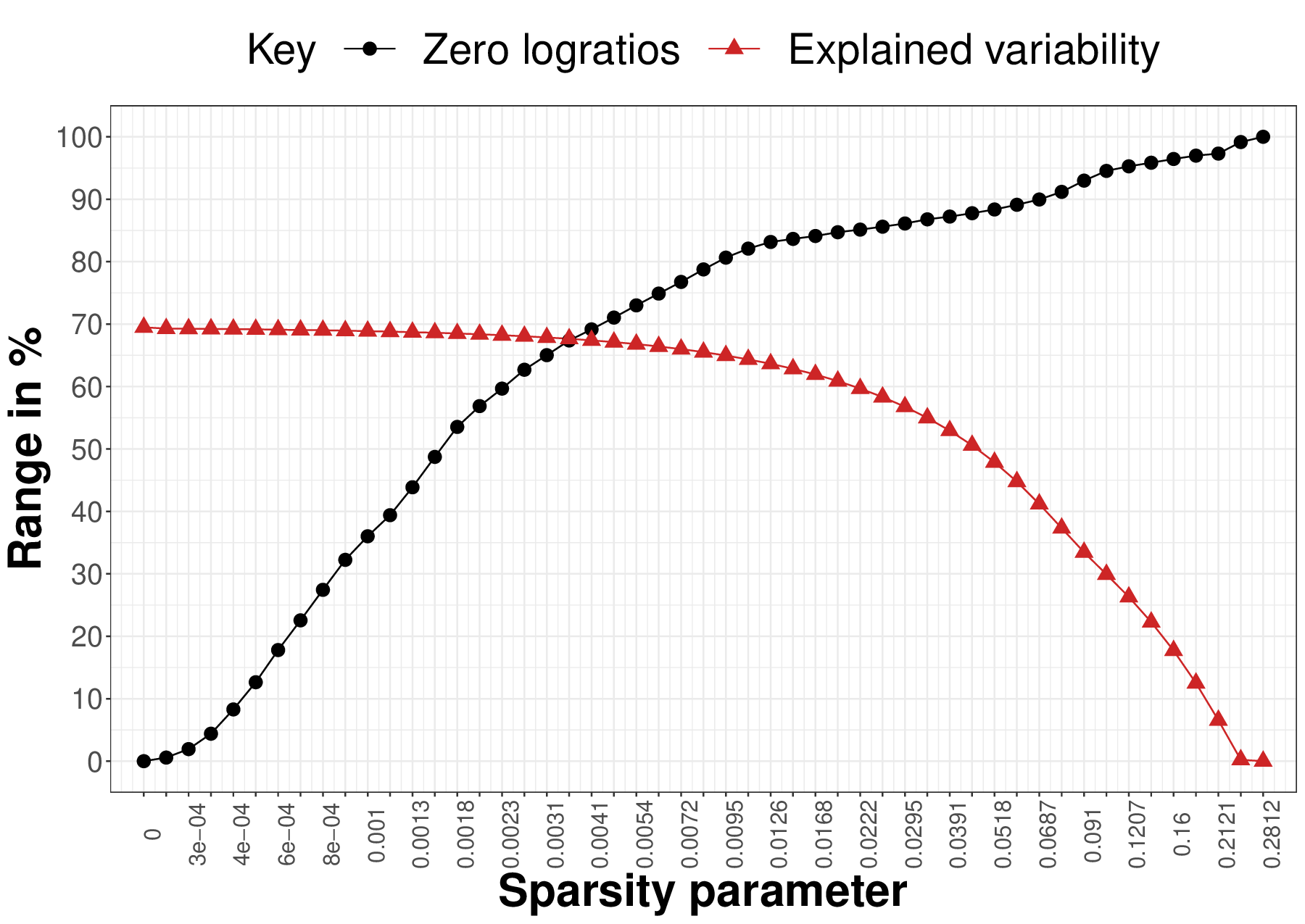}
\caption{Development of sparsity with increasing $\alpha$}
\label{subfig:2a}
\end{subfigure}%
\begin{subfigure}{0.475\textwidth}    
\centering
\includegraphics[width=\textwidth]{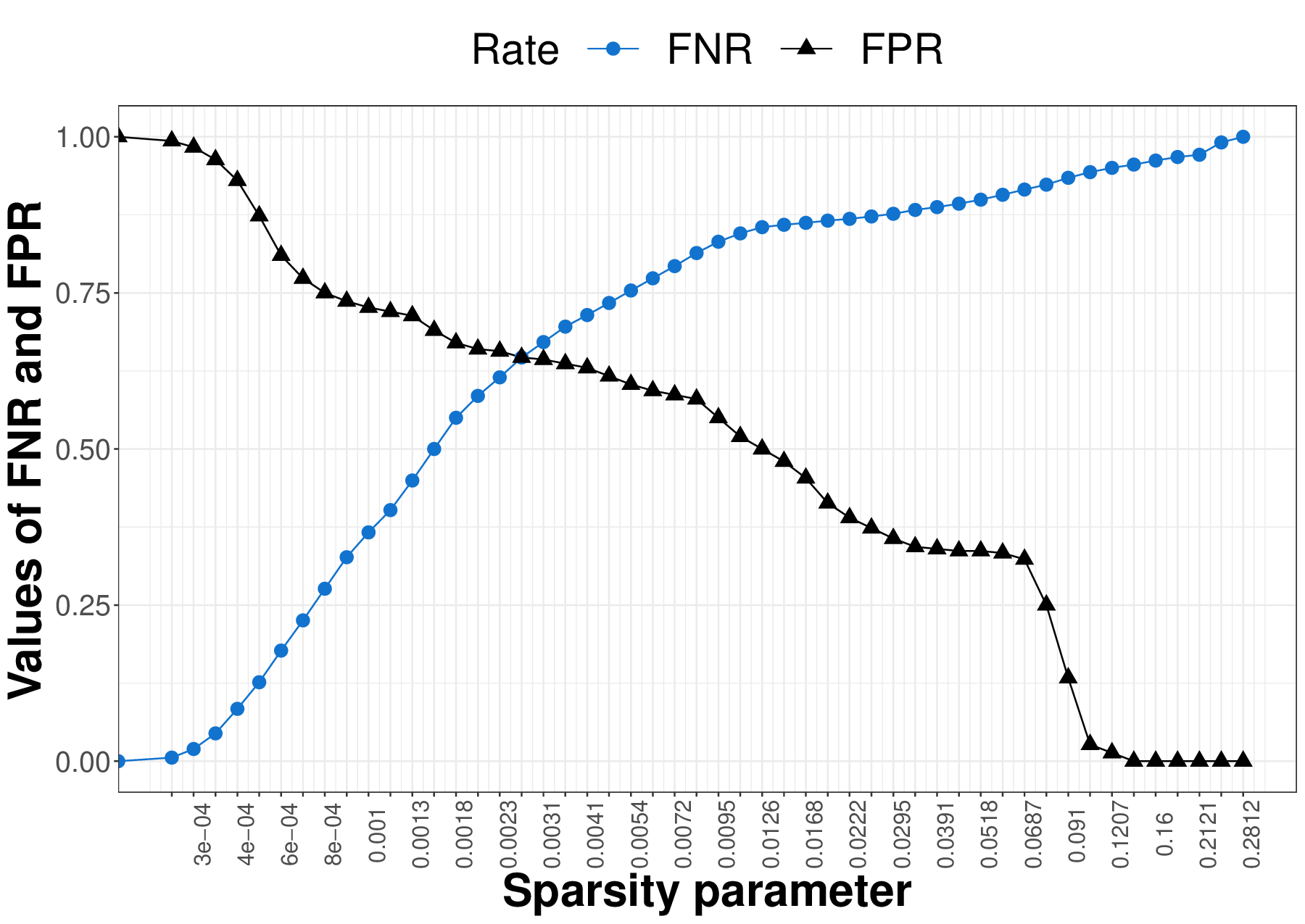}
\caption{FNR and FPR with increasing $\alpha$}
\label{subfig:2b}
\end{subfigure}
\caption{Simulation scenario B for $D=10$. Left: percentage of zero logratios (black) and explained variability (red) for different values of the sparsity parameter $\alpha$. Right: FPR (black) and FNR (blue) for different values of $\alpha$. }
\label{fig:SimB}
\end{figure}

{\bf Results Scenario B.} Results for scenario B are shown in Fig. \ref{fig:SimB}. 
Having more relevant balances leads to an overall higher percentage of explained variability (roughly 70\%) for the dense solution.
Similarly to scenario A, the percentage of explained variability decreases only slowly for sparser solutions but the sparsity of the loadings shoots up more quickly.
For a sparse solution where roughly half of the logratios have zero loadings, still almost 69\% of the variability can be explained. 
In this model, the $FNR=0.5$, which is considerably below the $FPR\mbox{ } (0.69)$, 
meaning that fewer important logratios are  missed (impacting the FNR) than that unimportant logratios are selected (impacting the FPR).

\begin{figure}
\centering
\begin{subfigure}{0.475\textwidth}   
\centering
\includegraphics[width=\textwidth]{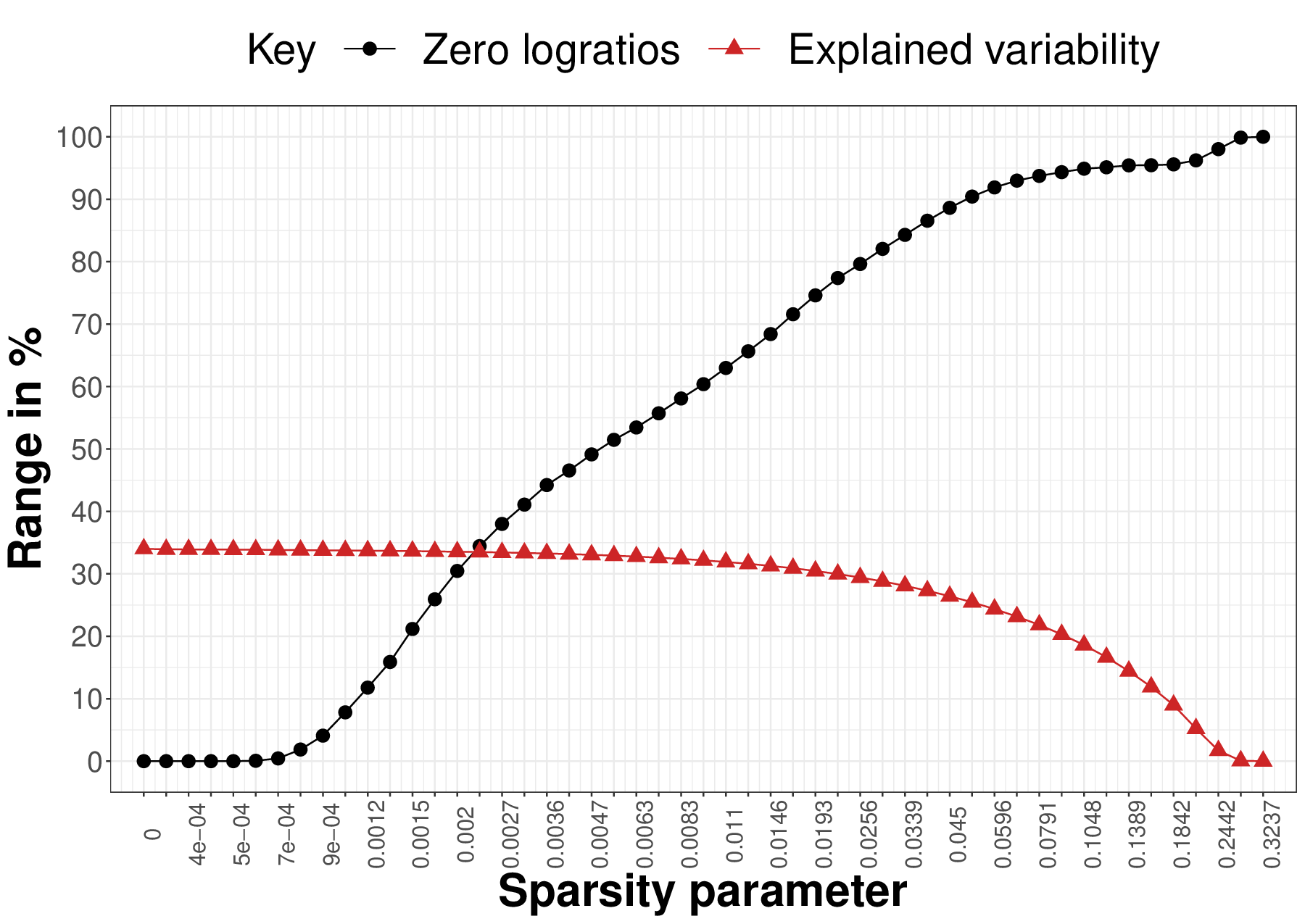}
\caption{Development of sparsity with increasing $\alpha$}
\label{subfig:3a}
\end{subfigure}%
\begin{subfigure}{0.475\textwidth}    
\centering
\includegraphics[width=\textwidth]{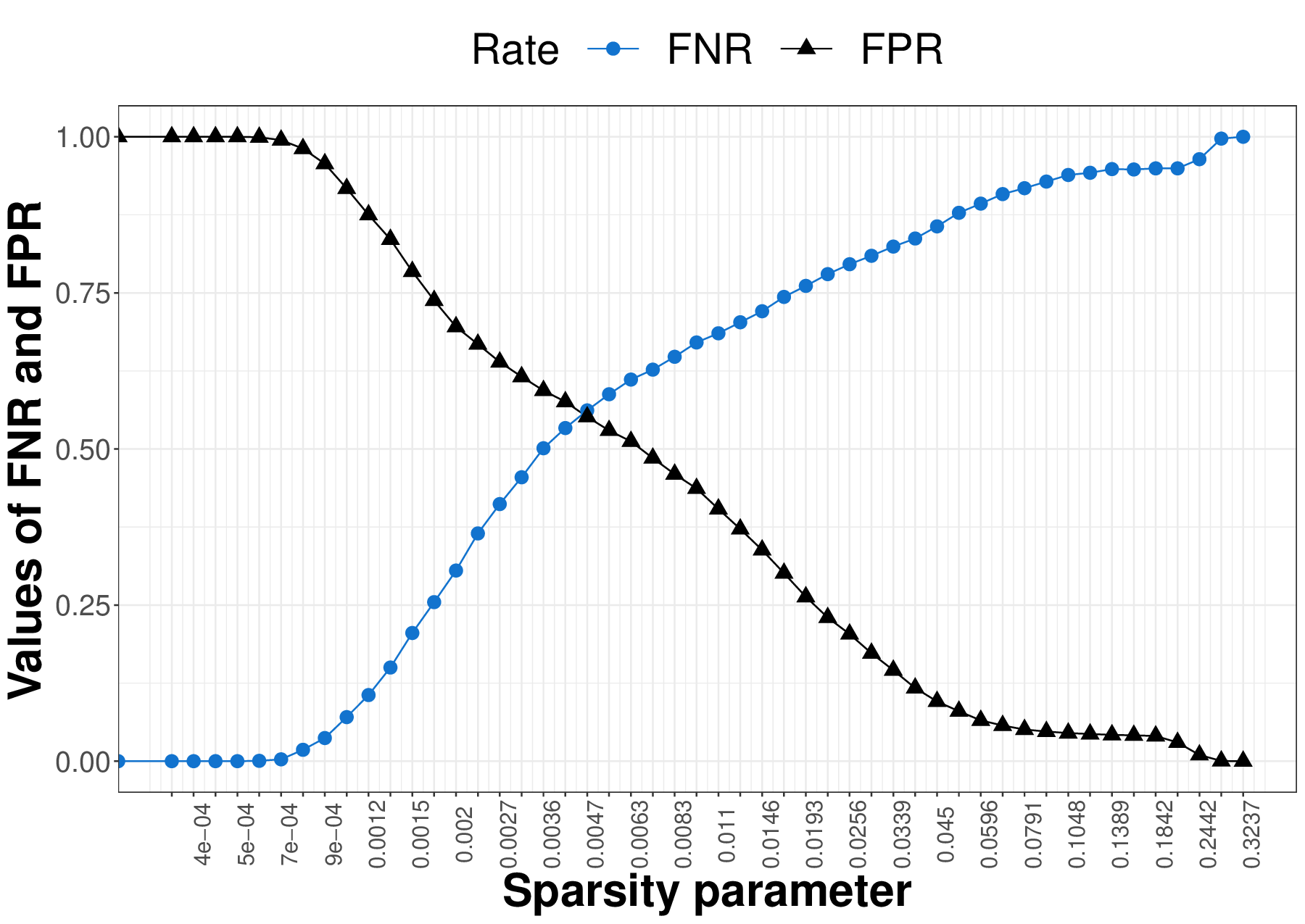}
\caption{FNR and FPR with increasing $\alpha$}
\label{subfig:3b}
\end{subfigure}
\caption{Simulation scenario C for $D=10$. Left: percentage of zero logratios (black)  and explained variability (red) for different values of the sparsity parameter $\alpha$. Right: FPR (black) and FNR (blue) for different values of $\alpha$.}
\label{fig:SimC}
\end{figure}

{\bf Results Scenario C.} Figure \ref{fig:SimC} shows results for the last scenario having most noise balances. The maximum percentage of explained variability (at the dense model) now drops below $40$\% but we still see an only moderate decline in explained variability combined with a steeper rise in sparsity when the sparsity parameter $\alpha$ gradually moves away from zero. 
Finally, the rate of increase in FNR, tracking the omission of important logratios, is
slower in scenario C (Fig. \ref{subfig:3b}) than in scenario B (Fig. \ref{subfig:2b}). Hence, when there are only few relevant balances, the important logratios ``survive" (i.e.\ display non-zero loadings) for a longer range of increasing values of the sparsity parameter.

\section{Applications} \label{real}
In this section, we demonstrate the usefulness of the sparse PCA proposal on two empirical data sets in identifying potentially relevant PLRs (and thus important parts of compositions). Both data sets consist of geochemical elements. The first data set contains compositions of glaciar sediments, the second data set comes from archaeometry.

\subsection{Aar Massif Data}
The data set Aar, introduced in \cite{von2012sediment}, contains geochemical CoDa of glaciar sediments from the Aar Massif in Switzerland.
The data  is publicly available in package \texttt{compositions} \citep{compositionsLib} of R.
We analyze a composition consisting of $D=10$ oxide components,  namely $\textrm{SiO}_{2}$, $\textrm{Al}_{2}\textrm{O}_{3}$, $\textrm{TiO}_{2}$, $\textrm{Fe}_{2}\textrm{O}_{3}^{tot}$ (total iron 3 oxide, further noted as $3t$), MnO, MgO, CaO, $\textrm{Na}_{2}\textrm{O}$, $\textrm{K}_{2}\textrm{O}$ and $\textrm{P}_{2}\textrm{O}_{5}$.
There are $n=87$ observations (i.e. samples) available on this 10-part composition. 
Hence, the total number of PLRs is $D(D-1) = 90$. Finally, there are no zero values in the data, hence imputation of zeros is not necessary.
Similarly to Sect. \ref{Sim}, we use sparse PCA to compute the first two PCs for (51) different values of the sparsity parameter $\alpha$, ranging from zero to the value $\alpha_{\text{max}}$ resulting in full sparsity. 

\begin{figure}
\centering
\includegraphics[width=0.8\textwidth]{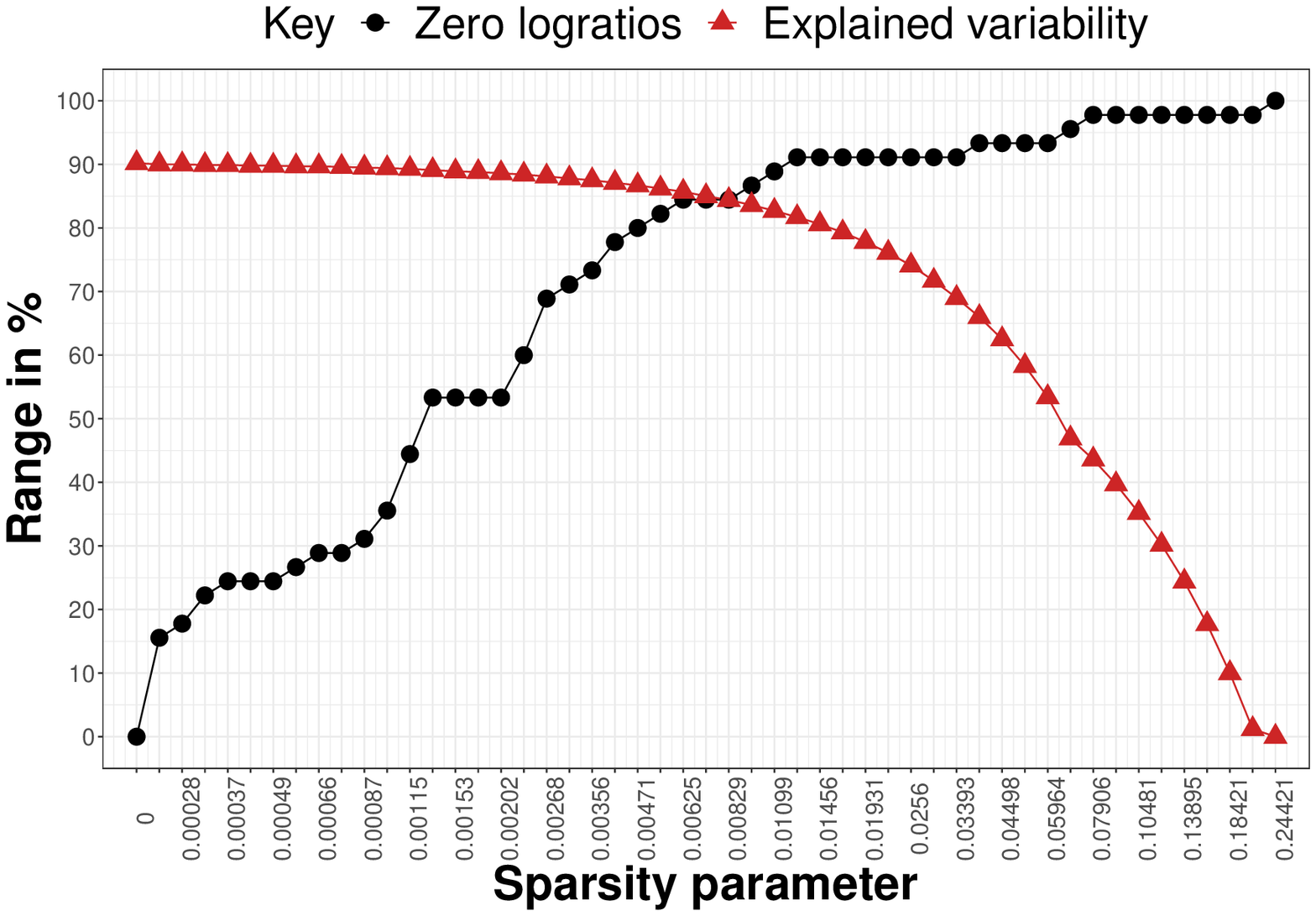}
  \caption{Aar Massif data: percentage of zero logratios (black) and  explained variability (red) by the first two PCs in the sparse PCA model with sparsity parameter $\alpha$ (on horizontal axis).
  }
  \label{fig:Aar}
\end{figure}

Figure \ref{fig:Aar} shows the trade-off between sparsity and explained variability for different values of the sparsity parameter $\alpha$.
The percentage of explained variability in the dense model is high, namely around 90\%. 
This percentage  declines only slowly and remains around 88\% even when more than half of the PLRs are set to zero (from $\alpha = 0.0013$ to $\alpha = 0.002$). Moreover, when the sparsity level is fairly high (almost 85\% of zero logratios, $\alpha = 0.0063$), the percentage of explained variability is still around 86\%. Sparse PCA solutions in this neighborhood still offer high percentages of explained variability (note that the percentage of explained variability drops below 80\% for models where slightly more than 90\% of PLRs were set to zero) and are thus good candidate solutions to identify important logratios.
Besides, from the simulation study, we know that the best balance in terms of FPR and FNR is typically reached for values of the sparsity parameter where the two lines (of explained variability and zero loadings) cross. This happens when there are more than 84\% of zero PLRs and $\alpha = 0.0072$ (or the two neighbouring values of $\alpha$, as for each we reach the same sparsity and the explained variability changes just slightly from 85.7\% to 84.4\%). 
Finally, note that the PCA solutions with the highest sparsity levels (i.e.\ solutions corresponding to the largest  $\alpha$ values) should be omitted from one's consideration as the (close to maximal) sparsity level makes the model meaningless due to the zero percentage of explained variability. 

\begin{figure}[t]
\includegraphics[width=1\textwidth]{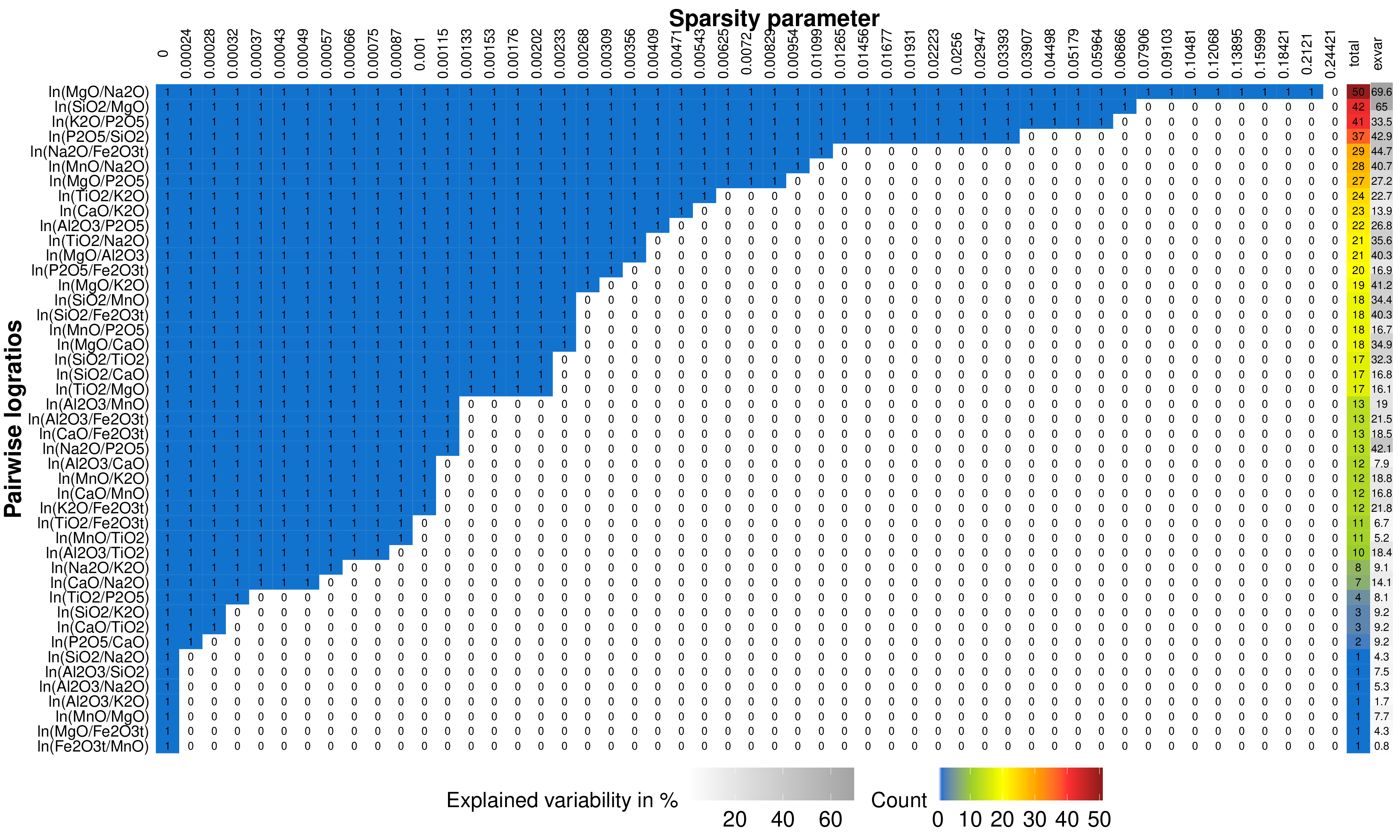}
  \caption{Aar Massif data: stability paths for PLRs (in rows) where coloured cells with ones reflect selection of the PLR (i.e.\ non-zero loading) for the sparse PCA solution with particular $\alpha$ (in columns). 
  The column ``total" displays the number of models (row counts) out of 51 in which the PLR is selected, the column ``exvar" gives the percentage of  variability explained by the logratio.}
  \label{fig:AarHeatWhole}
\end{figure}

Next, we investigate the stability selection of each PLR across the sparse PCA models (with varying sparsity parameter) through the stability paths depicted in Figure \ref{fig:AarHeatWhole}, which is similar to Fig. \ref{fig:Sim1Aheat} in the simulation study. 
From Fig. \ref{fig:AarHeatWhole}, we observe that the logratios $\textrm{ln}\frac{MgO}{Na_{2}O}$, $\textrm{ln}\frac{SiO_{2}}{MgO}$, $\textrm{ln}\frac{K_{2}O}{P_{2}O_{5}}$ and $\textrm{ln}\frac{P_{2}O_{5}}{SiO_{2}}$ are very stable as they are selected across more than 25 considered models (with $\textrm{ln}\frac{MgO}{Na_{2}O}$ being the most stable one), which can be seen from the row counts in the before last column of the figure. 
Hence, these PLRs are to be considered as important.
In the last column of Fig. \ref{fig:AarHeatWhole}, we display the contribution of each PLR to the total explained variability. 
We see that the most stable PLRs also display among the largest contributions to the explained variability, ranging from almost 70\% to roughly 34\%.
At the other end of the spectrum, our sparse PCA procedure returns zero loadings for altogether seven PLRs across practically all models except the dense one, thereby deeming these as rather unimportant. 
However, these logratios differ 
in the explained variability, ranging from around 7.5\% for $\textrm{ln}\frac{Al_{2}O_{3}}{SiO_{2}}$ to roughly 0.8\% for $\textrm{ln}\frac{Fe_{2}O_{3t}}{MnO}$. Except for $\textrm{ln}\frac{SiO_{2}}{MgO}$,  \cite{greenacre2019variable} identified the aforementioned most stable logratios as those mostly contributing to the variance.

\begin{figure}
\centering
\includegraphics[width=1\textwidth]{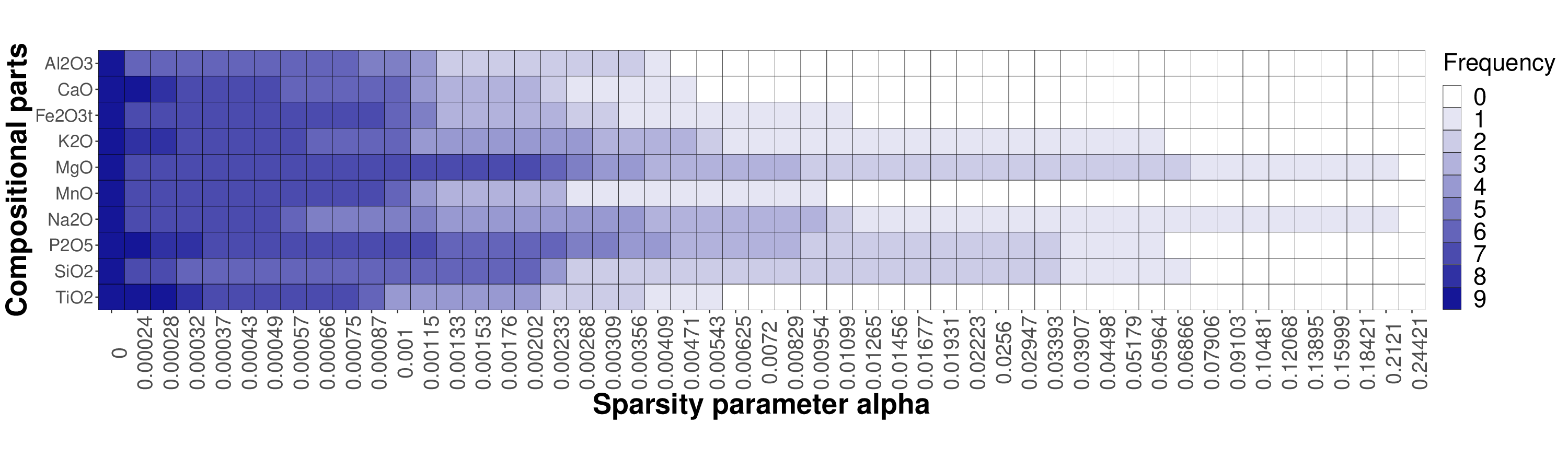}
  \caption{Aar Massif data: heatmap of compositional parts (in rows) for the sparse PCA solution with particular $\alpha$ (in columns). The darker the shading, the higher the occurrence of the part across non-zero PLRs.}
  \label{fig:AarHeatParts}
\end{figure}

Next, we further zoom out to the compositional parts and compute their frequency of occurrence and stability across the sparse PCA models.
In Fig. \ref{fig:AarHeatParts}, we now display the 10 components on the vertical axis, while keeping the sparsity parameter on the horizontal axis. Each cell then displays the number of non-zero PLRs in which the particular part appears.
The darker the cell, the more frequent the part occurs.
Besides, the further the colored shading stretches to the right of the plot, the higher the stability of the part.
We observe that MgO, $\textrm{P}_{2}\textrm{O}_{5}$ and $\textrm{SiO}_{2}$ are most frequently selected in the first seventeen or eighteen models. 
From another perspective, MgO, $\textrm{Na}_{2}\textrm{O}$ (followed by $\textrm{K}_{2}\textrm{O}$, $\textrm{P}_{2}\textrm{O}_{5}$ and $\textrm{SiO}_{2}$) are most stable as they are present in the majority of models, even though not so frequently. This is not surprising as the PLRs with MgO, $\textrm{Na}_{2}\textrm{O}$,  $\textrm{K}_{2}\textrm{O}$, $\textrm{P}_{2}\textrm{O}_{5}$ and $\textrm{SiO}_{2}$ turned out to be within the most stable ones (see Fig. \ref{fig:AarHeatWhole}).
Besides, the elements Mg, Na, Si and P are documented to have the largest contribution to the explained variability in \cite{von2012sediment}. We also find these to play a prominent role in the heat map of Fig. \ref{fig:AarHeatParts}, most of them mainly in terms of stability but Mg, P and Si (in parts MgO, $P_2O_5$ and $\textrm{SiO}_{2}$) also in terms of frequency of occurrence. 

\subsection{Archaeometric Data}
Our second data set contains samples of Roman glass cups found in an archaeological site in eastern England. The data are introduced in \cite{baxter1990principal} and also analyzed in \cite{greenacre2019variable}. We have a $D=11$-part composition of elements silicon (Si), aluminium (Al), iron (Fe), magnesium (Mg), calcium (Ca), sodium (Na), potassium (K), titanium (Ti), phosphorus (P), manganese (Mn) and antimony (Sb) for which $n=47$ observations are available. There are no zeros in the data, hence imputation is not needed.

\begin{figure}
\centering
\includegraphics[width=0.8\textwidth]{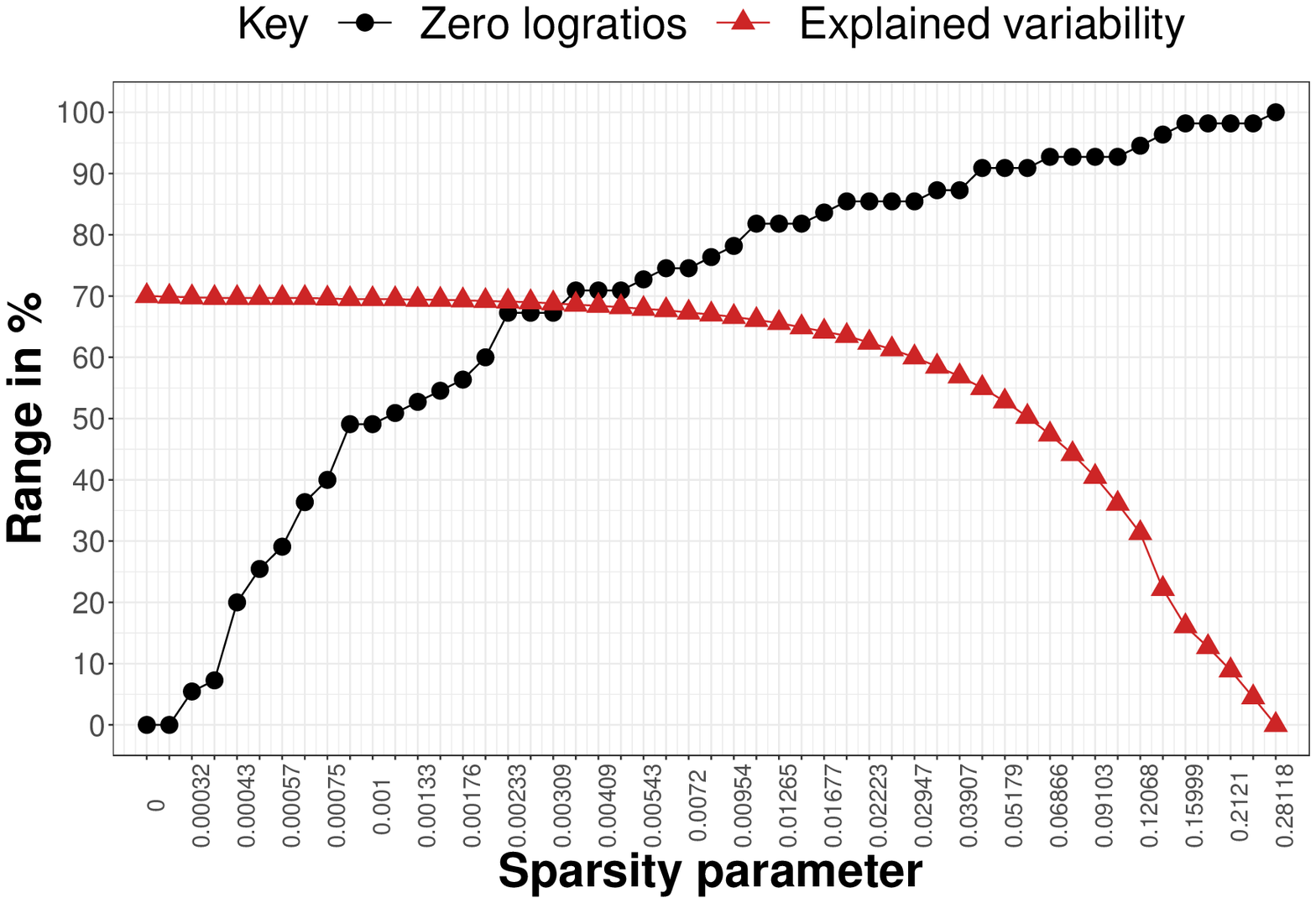}
  \caption{Archaeometric data: percentage of zero logratios (black) and  explained variability (red) by the first two PCs in the sparse PCA model with sparsity parameter $\alpha$ (on horizontal axis).
  }
  \label{fig:ArchLine}
\end{figure}

We conduct the same analysis as for the Aar Massif data set. Figure \ref{fig:ArchLine} gives the trade-off between sparsity and explained variability across the various sparse PCA solutions. It can be observed that sparsity
for $\alpha = 0.001$ results in a model with roughly 50\% of zero PLRs while still reaching almost 70\% of explained variability. Interestingly, the percentages of explained variability remains considerably high and, importantly, flat at the start of the plot. Even for more than $60$\% zero PLRs, the explained variability remains still almost $70$\%. This data set therefore seems to lean itself well for sparsification.

\begin{figure}
\includegraphics[width=\textwidth]{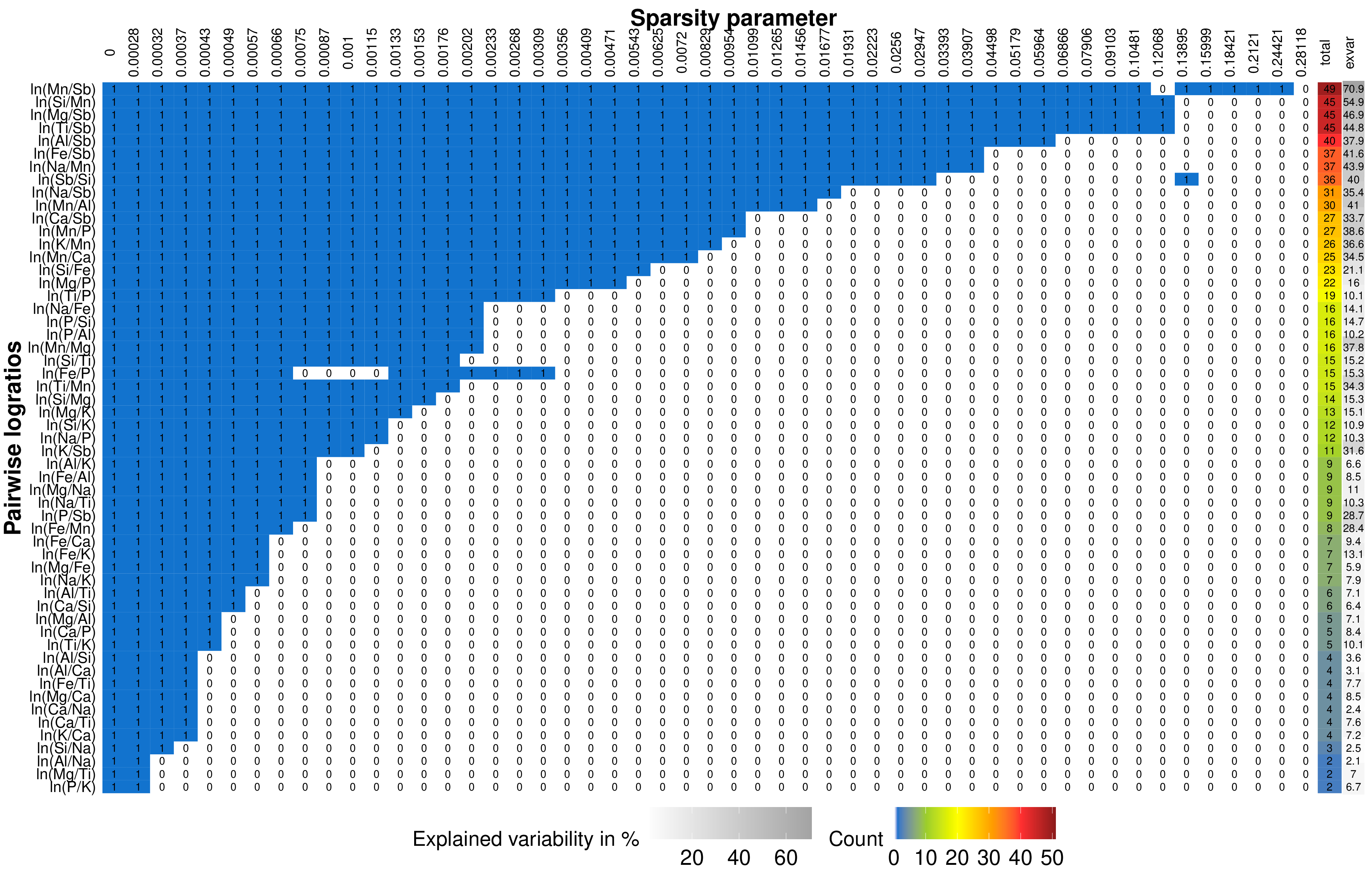}
  \caption{Archaeometric data: stability paths for PLRs (in rows) where coloured cells with ones reflect selection of the PLR (i.e.\ non-zero loading) for the sparse PCA solution with particular $\alpha$ (in columns). 
  The column ``total" displays the number of models (row counts) out of 51 in which the PLR is selected, the column ``exvar" gives the percentage of  variability explained by the  logratio.}
  \label{fig:Gre_stab}
\end{figure}

Next, we inspect the stability of the PLRs in  Fig. \ref{fig:Gre_stab}. The most stable PLRs are $\textrm{ln}\frac{Mn}{Sb}$, $\textrm{ln}\frac{Si}{Mn}$, $\textrm{ln}\frac{Mg}{Sb}$ and $\textrm{ln}\frac{Ti}{Sb}$ with explained variability of $70.85$\%, $54.88$\%, $46.91\%$ and $44.83\%$ respectively. These logratios appeared to be non-zero in more than 45 models. One can observe a slightly odd ``gap`` for $\alpha = 0.12068$ considering the first logratio and three other cells which do not follow the assumption of continuously increasing sparsity. This could be result of certain slight unstability of the procedure in presence of complex data structure, however still these effects are very minor. Apart from the above mentioned four most stable logratios, there is also another group of four logratios having non-zero loadings across more than 35 (i.e. more than half, having altogether 51 models) sparse PCA solutions.
When considering the sparsest model including all only the four most stable logratios logratios (for $\alpha=0.10481$, as in the next model the first logratio is set as zero), still up to roughly 36\% of the total variability is explained (as can be seen from the red curve with corresponding $\alpha$ value in Fig. \ref{fig:ArchLine}).  \cite{greenacre2019variable} identifies the three  PLRs, namely $\textrm{ln}\frac{Si}{Ca}$, $\textrm{ln}\frac{Si}{Sb}$ and $\textrm{ln}\frac{Na}{Sb}$ as being most important using a stepwise analysis. The latter two also appear to be considerably stable in our analysis as they appear as non-zero in more than 30 models.

\begin{figure}
\includegraphics[width=\textwidth]{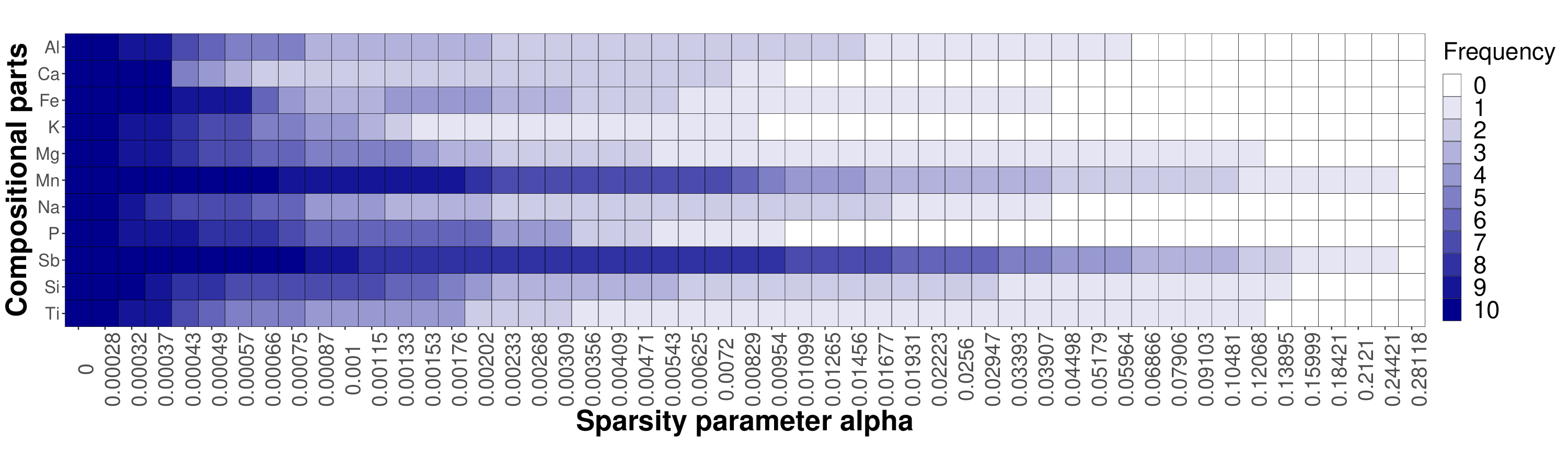}
  \caption{Archaeometric data: heatmap of compositional parts (in rows) for the sparse PCA solution with particular $\alpha$ (in columns). The darker the shading, the higher the occurrence of the part across non-zero PLRs.}
  \label{fig:ArchParts}
\end{figure}

Finally, we consider the heatmap for compositional parts in Fig. \ref{fig:ArchParts}. 
Especially elements Mn and Sb are most frequently occurring across selected logratios. 
Moreover, these parts also ``survive" the longest, as their coloring stretches most to the right of the plot. It can be seen that also elements Mg, Si and Ti appear in non-zero logratios even in highly sparse models, even though not so frequently as Mn and Sb.
Besides, all parts occur in at least one selected PLR even in sparse PCA solutions.

\section{Summary and conclusions}  \label{End}

Compositional data analysis is based on logratio information of the 
variables. The simplest form of a logratio is in terms of the logarithm of a
ratio of two variables, and this is also simple for the interpretation. The goal
of this paper thus was to identify pairwise logratios (PLRs) which together
explain the most relevant information contained in the compositional data set.
In order to identify those PLRs, we used a sparse PCA method applied on
the matrix of all possible (relevant) PLRs. Due to the sparsity of the method, some or many of the 
loading entries are zero, and only the non-zero entries are pointing at
important PLRs. The number of non-zero entries is determined by a sparsity
parameter, and we presented a visualization where the tradeoff between
sparsity and explained variance can be balanced. In our examples,
only the first two principal components were used; in general, for more
complex data, one could select the number of components in order to explain
around 80\% of the variance. The same plot can then still be used to find 
the appropriate compromise between reducing explained variance and increasing sparsity.

We presented simulation studies where the true important logratios were
known. The proposed methodology turned out to be successful in identifying
those PLRs. We compared with the algorithm STEP \citep{greenacre2019variable},
which has the goal to represent the data set in a new coordinate system
(without dimension reduction) which is built upon PLRs. Although the goal is thus somewhat different, our procedure is more successful in identifying
the true important PLRs.

In our empirical applications, we present several graphical tools to aid practitioners in their multivariate analysis with CoDa.
Next to the already mentioned plot to select the sparsity parameter,
we provide a visualization showing
stability paths for the PLRs. They reveal which PLRs 
remain in the PCA model if the sparsity of the PCA solution gradually increases.
The important, stable logratios highlighted by our analysis also turned out to be important in previous CoDa analysis on the considered geochemical data sets.
Finally, we provide heatmaps for the original compositional parts, thereby permitting to evaluate their importance for the multivariate analysis.

The paper thus shows a clear way how to get oriented among the usage of many PLRs in PCA, even for compositions with a moderate number of components. As the recent trend in CoDa analysis is to prefer simple PLRs over more sophisticated logratio coordinate representations, the use of sparse models with few PLRs promises to be of increasing importance also beyond the scope of PCA to, for instance, classification tasks.

\section*{Acknowledgments}
We acknowledge support through the HiTEc Cost Action
CA21163.
VN and KH gratefully acknowledge the support of the grant IGA\_PrF\_2022\_008 Mathematical models. PF acknowledges support
from the Austrian Science Fund (FWF), project number I 5799-N. 
KH was also supported by the Czech Science
Foundation grant 22-15684L and the project PID2021-123833OB-I00 provided by the Spanish Ministry of Science and Innovation (MCIN/AEI/10:13039/501100011033) and ERDF A way of making Europe.

\section*{Declarations}

{\textbf{Conflict of interest}} The authors declare no conflict of interest.

\bibliographystyle{apalike}
\bibliography{Main}
\nocite{*}

%

\newpage
\appendix
\section{Results for the different simulation scenarios with $D = 20$} \label{appA}

The following section shows simulation results for all three simulation scenarios A, B and C with $D = 20$ parts. The data are simulated using the procedure described in Section \ref{Sim}, considering 10 (scenario A), 15 (scenario B) and 4 (scenario C) relevant balances.

\begin{figure}[h]
\centering

\begin{subfigure}{0.475\textwidth}    
\centering
\includegraphics[width=\textwidth]{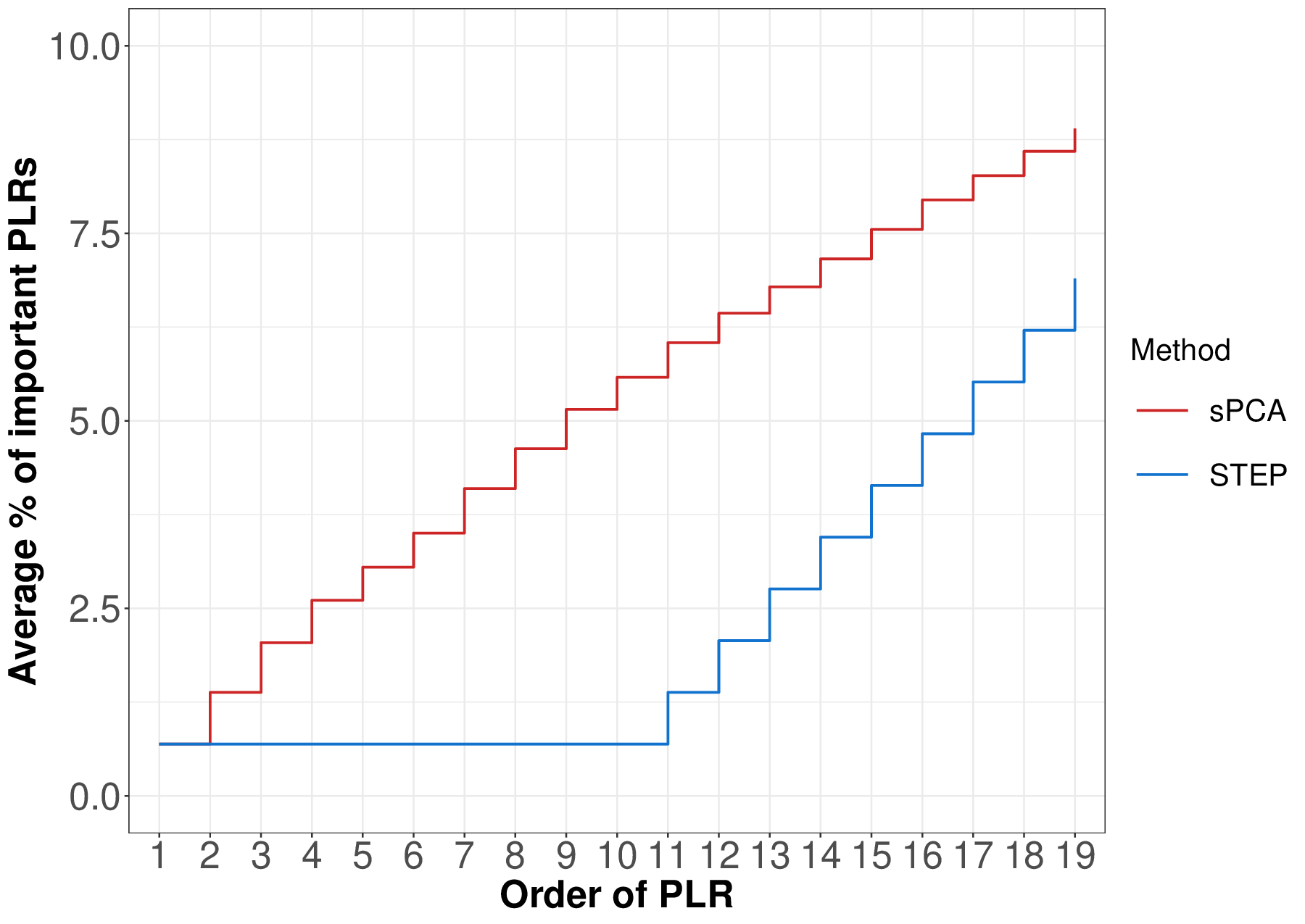}
\caption{Scenario A}
\label{2subfig:2b}
\end{subfigure}
\begin{subfigure}{0.475\textwidth}   
\centering
\includegraphics[width=\textwidth]{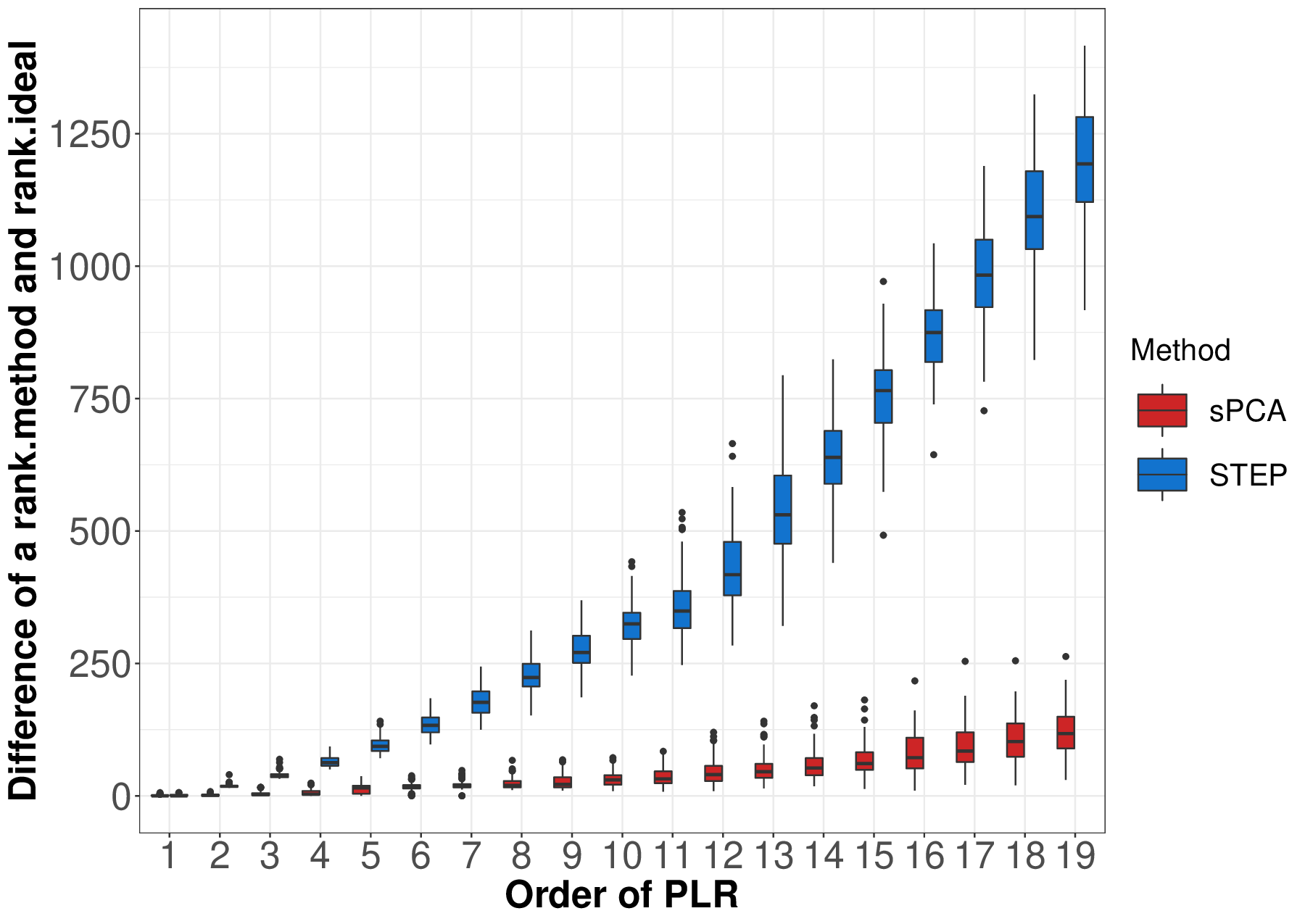}
\caption{Scenario A}
\label{2subfig:2a}
\end{subfigure}%

\begin{subfigure}{0.475\textwidth}    
\centering
\includegraphics[width=\textwidth]{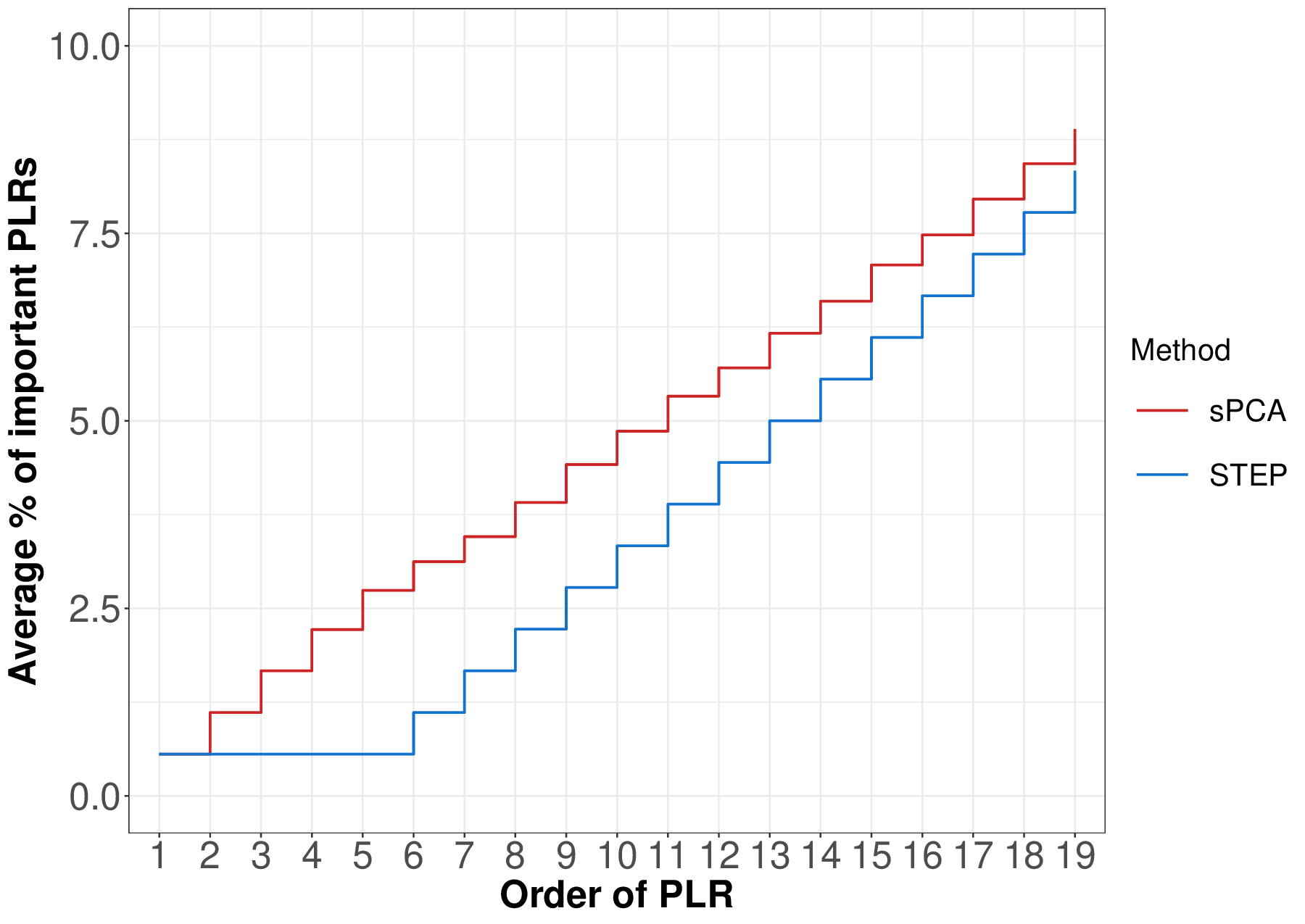}
\caption{Scenario B}
\label{2subfig:2d}
\end{subfigure}
\begin{subfigure}{0.475\textwidth}    
\centering
\includegraphics[width=\textwidth]{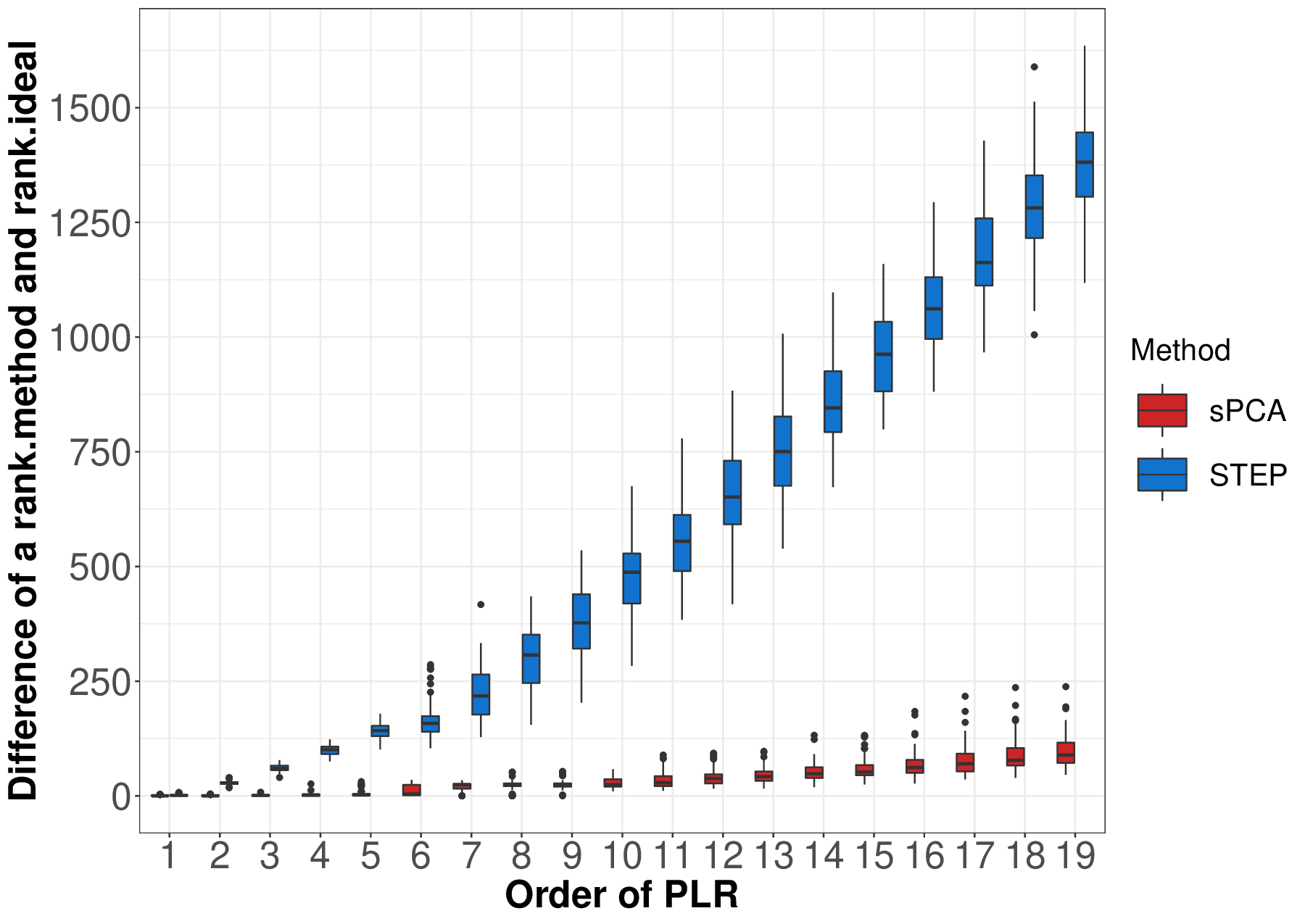}
\caption{Scenario B}
\label{2subfig:2c}
\end{subfigure}

\begin{subfigure}{0.475\textwidth}   
\centering
\includegraphics[width=\textwidth]{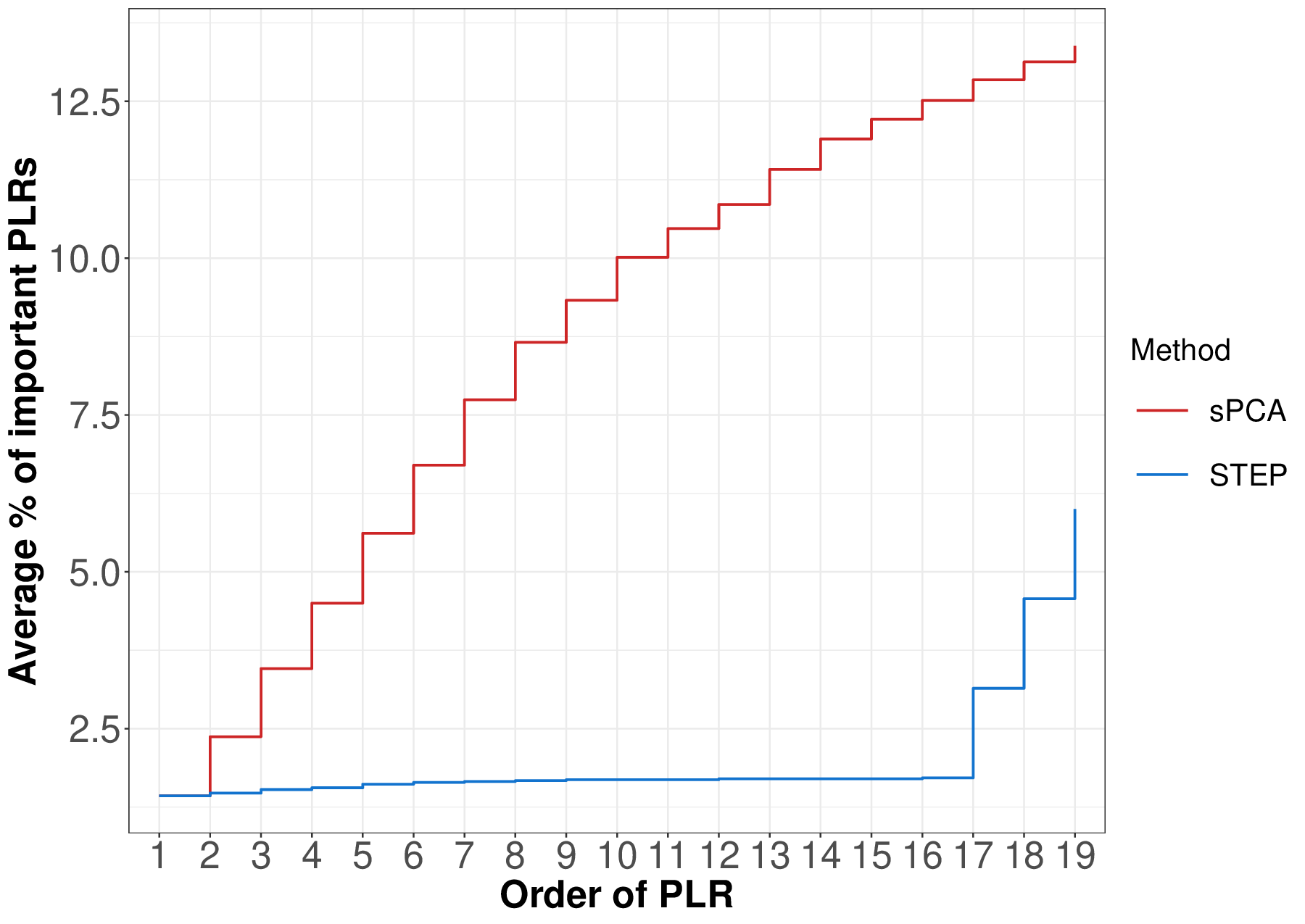}
\caption{Scenario C}
\label{2subfig:2f}
\end{subfigure}%
\begin{subfigure}{0.475\textwidth}   
\centering
\includegraphics[width=\textwidth]{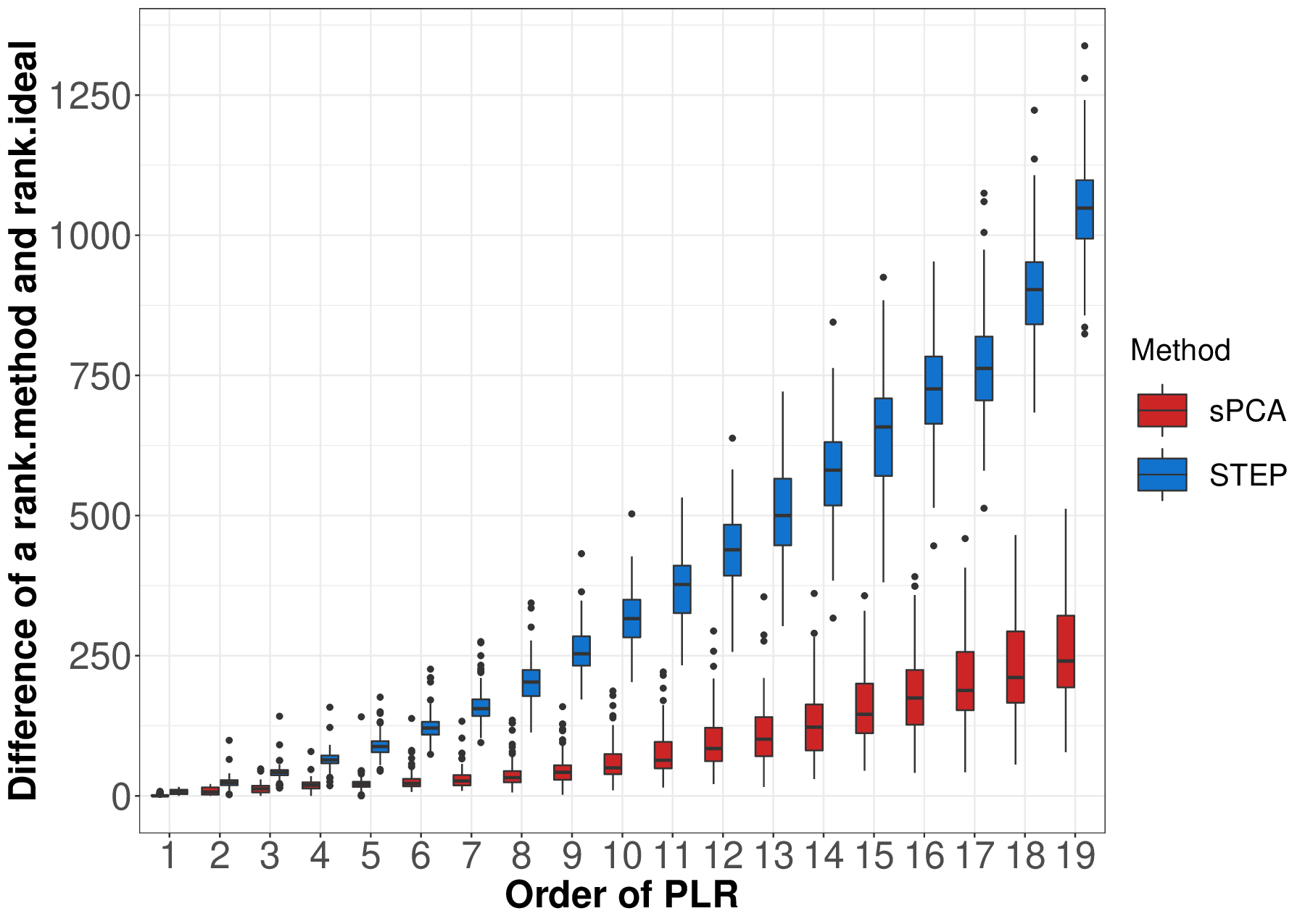}
\caption{Scenario C}
\label{2subfig:2e}
\end{subfigure}%

\caption{Comparison of sparse PCA to STEP on simulation scenarios A, B and C for $D=20$. 
Left:  average percentage of correctly identified PLRs for sPCA (red) and STEP (blue) for the first $D-1$ ordered PLRs (PLR) on the horizontal axis.
Right: boxplots showing the differences between the cumulative ranks of the PLR obtained from sPCA (red) or STEP (blue) and the ideal ranks. }
\label{fig:LogrsP2}
\end{figure}

When comparing sparse PCA and STEP via (i) the percentage of important PLRs and (ii) the rank of the chosen logratios, it can be seen in Fig. \ref{fig:LogrsP2} that sparse PCA  outperforms STEP on both criteria. Unlike the results for $D=10$, the advantage of sparse PCA over STEP is now also more apparent in simulation scenario C.

\begin{figure}[h]
\centering
\begin{subfigure}{0.475\textwidth}   
\centering
\includegraphics[width=\textwidth]{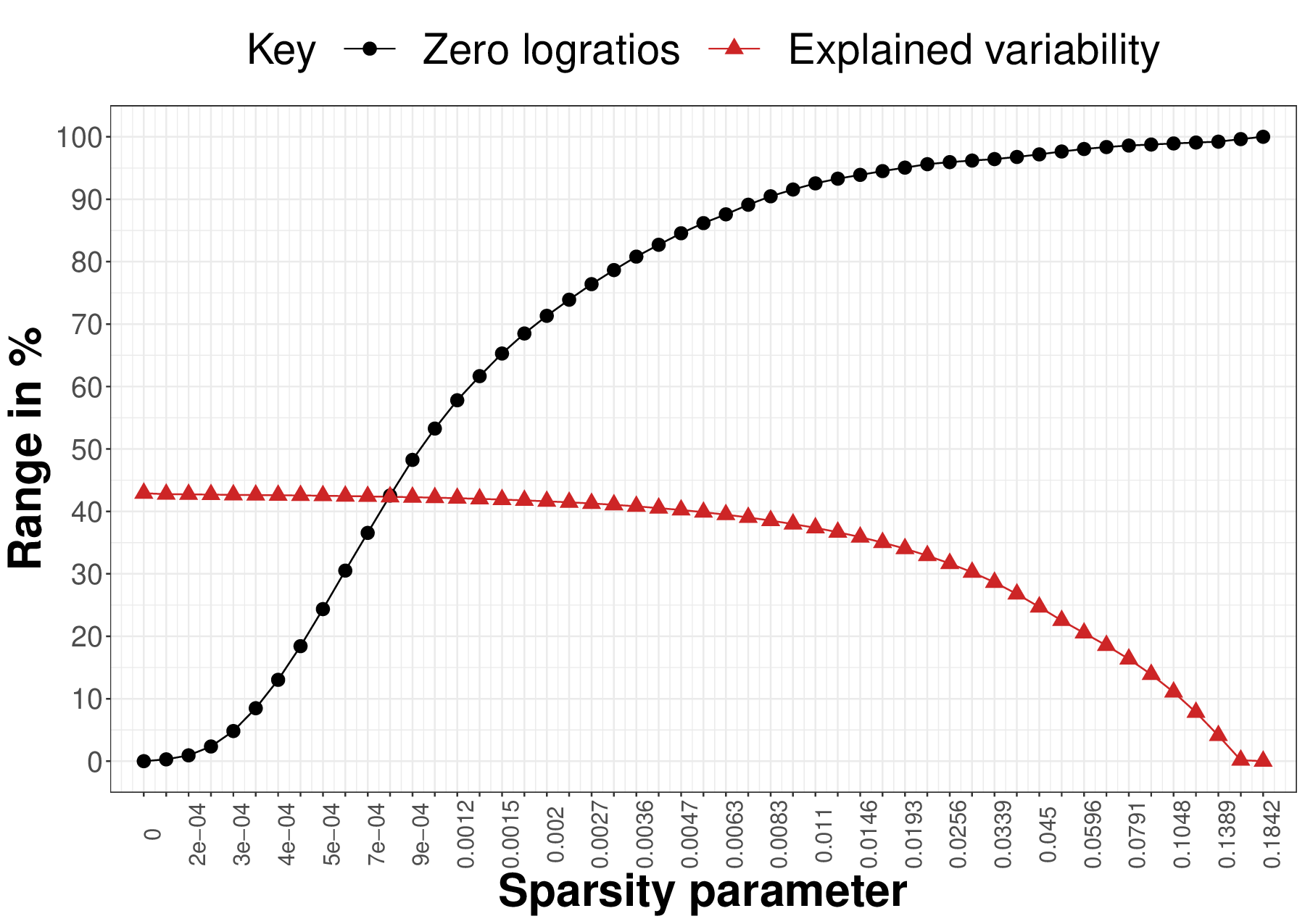}
\caption{Scenario A}
\label{3subfig:2a}
\end{subfigure}%
\begin{subfigure}{0.475\textwidth}    
\centering
\includegraphics[width=\textwidth]{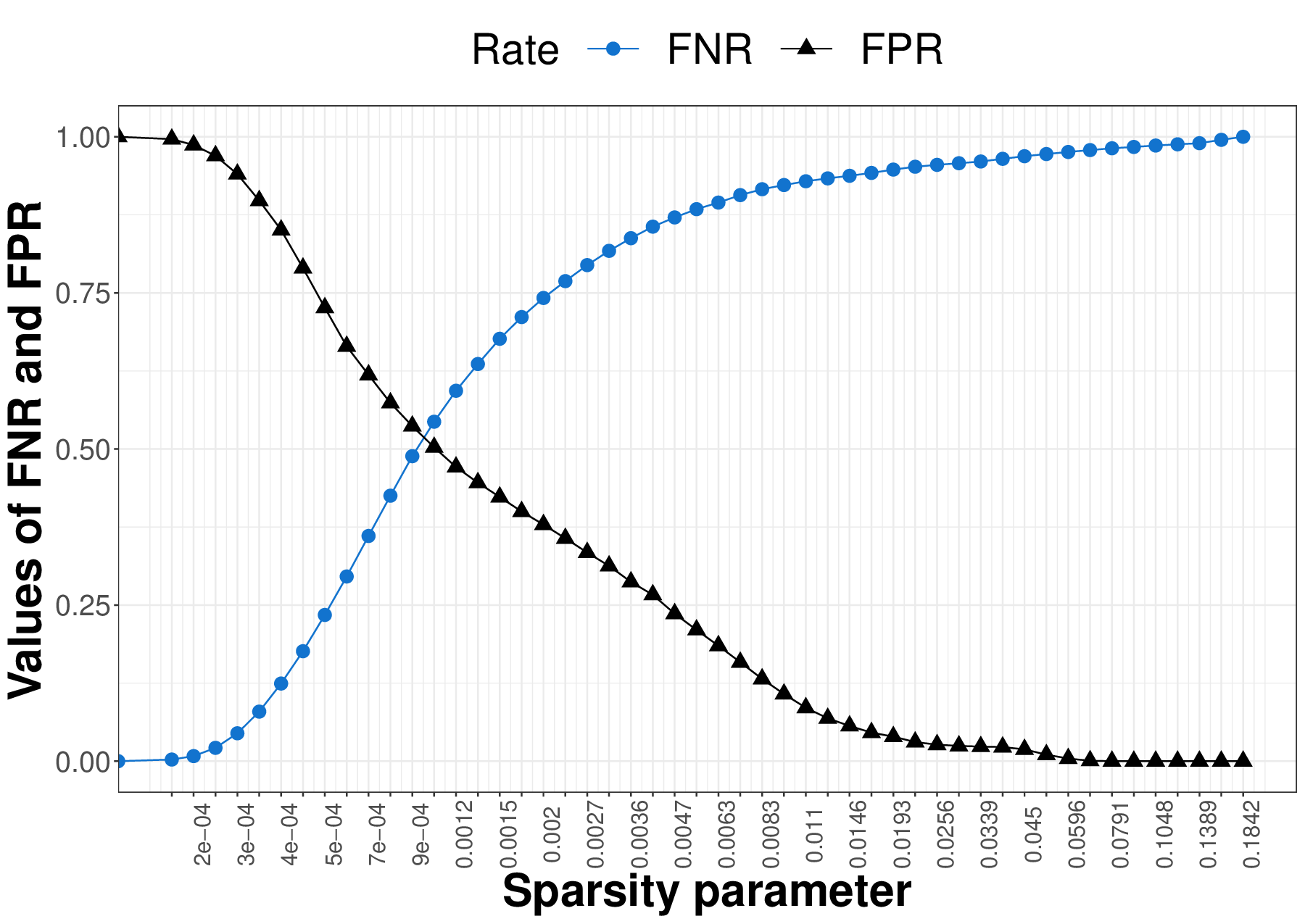}
\caption{Scenario A}
\label{3subfig:2b}
\end{subfigure}
\begin{subfigure}{0.475\textwidth}    
\centering
\includegraphics[width=\textwidth]{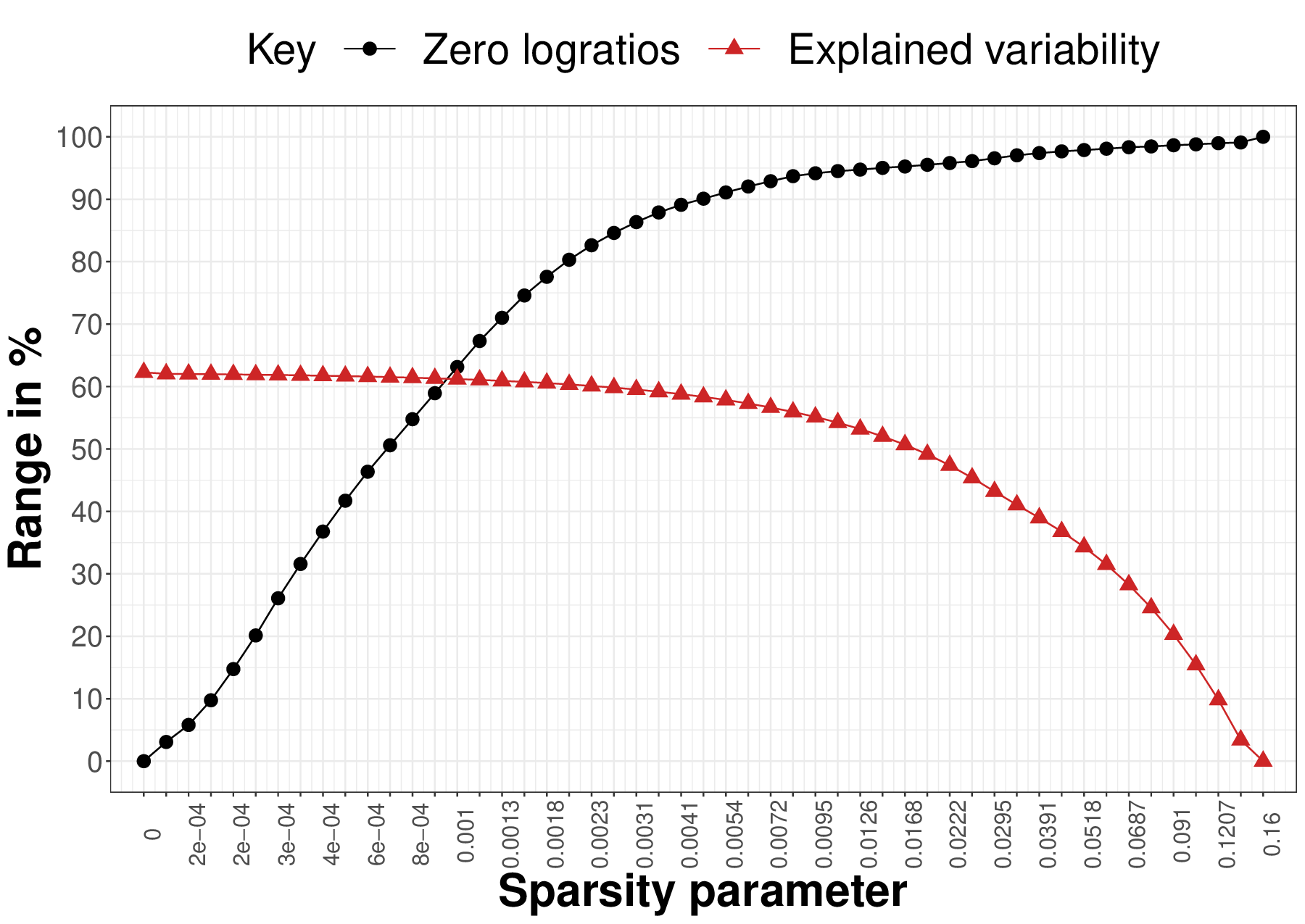}
\caption{Scenario B}
\label{3subfig:2c}
\end{subfigure}
\begin{subfigure}{0.475\textwidth}    
\centering
\includegraphics[width=\textwidth]{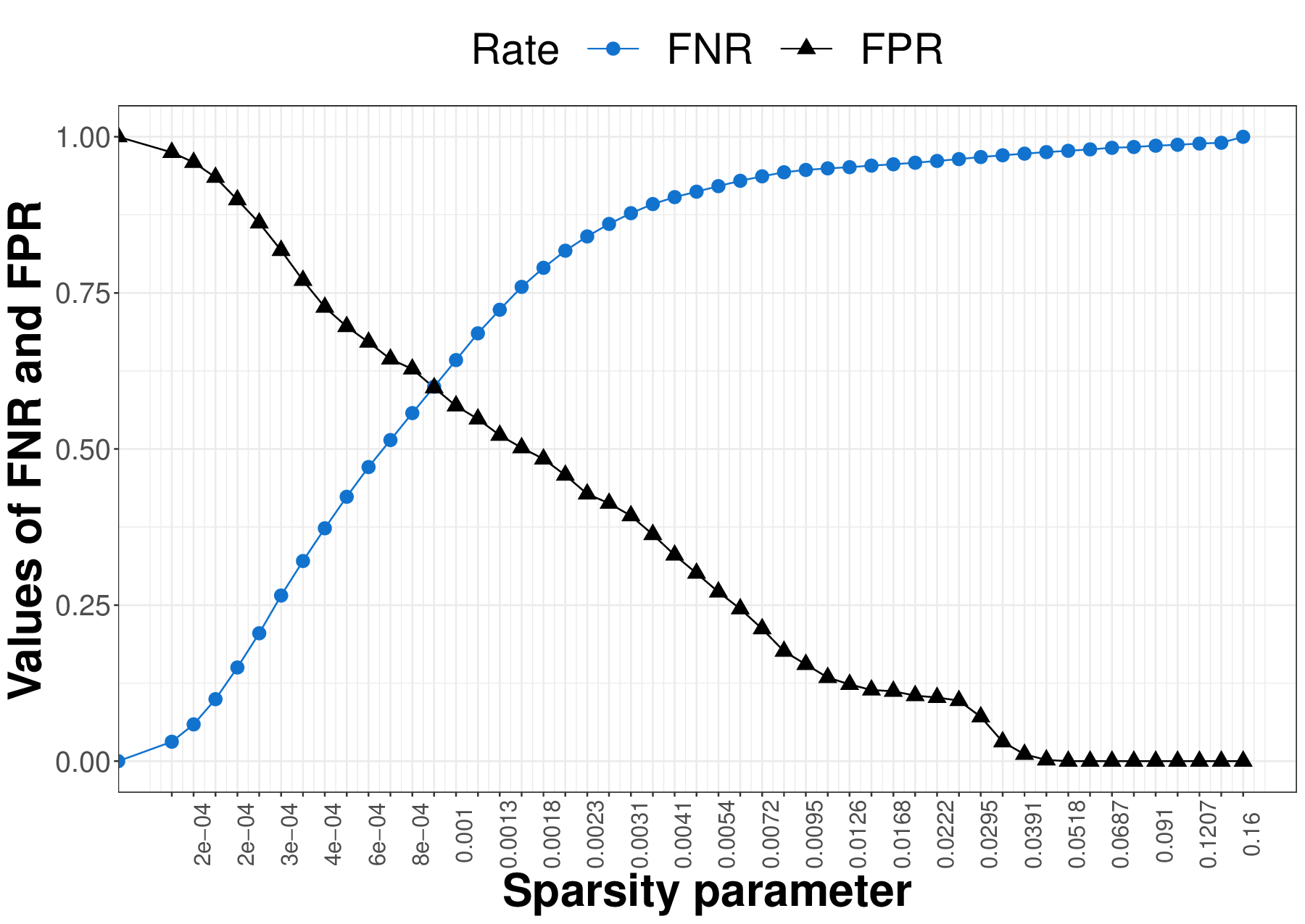}
\caption{Scenario B}
\label{3subfig:2d}
\end{subfigure}
\begin{subfigure}{0.475\textwidth}   
\centering
\includegraphics[width=\textwidth]{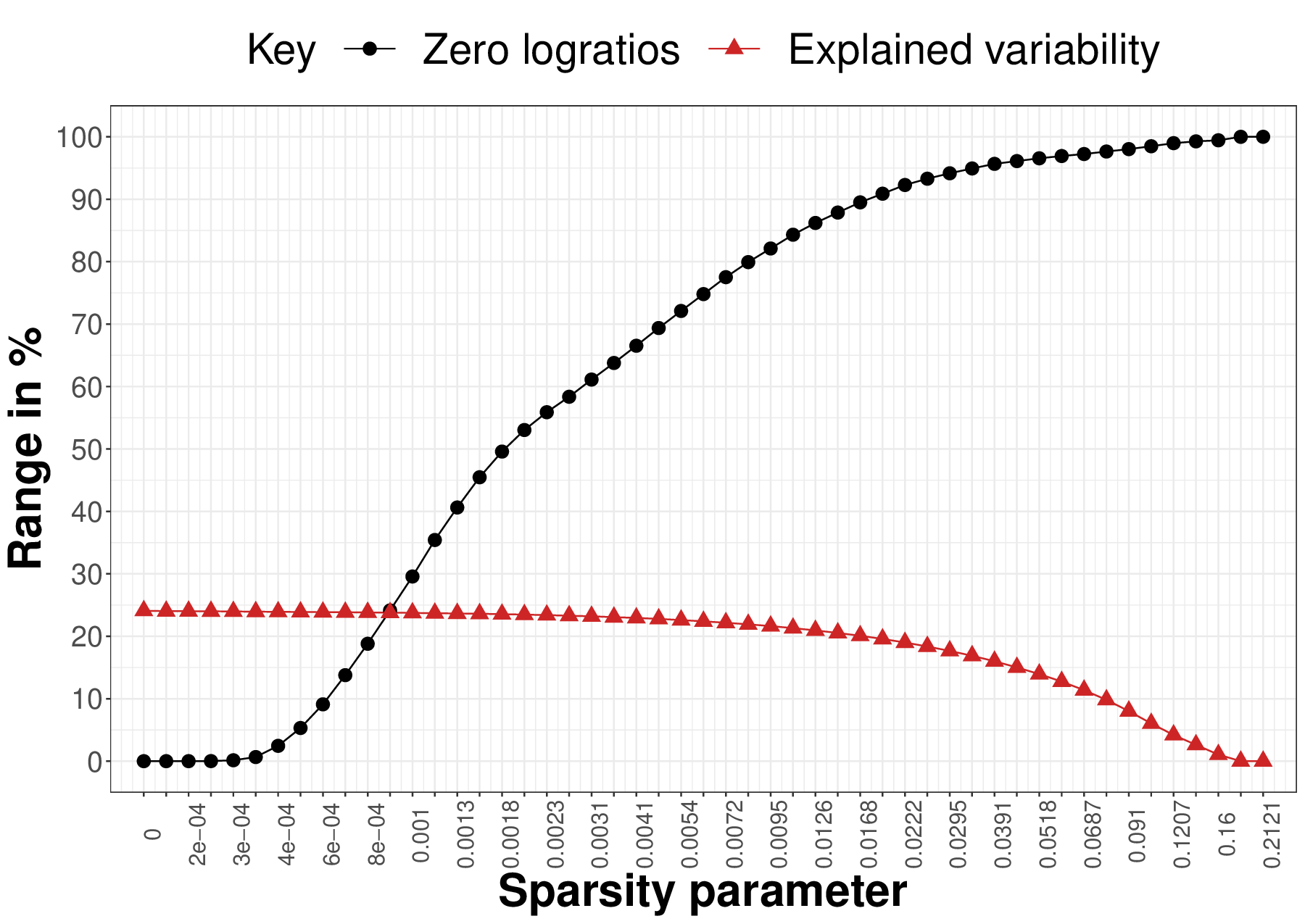}
\caption{Scenario C}
\label{3subfig:2e}
\end{subfigure}%
\begin{subfigure}{0.475\textwidth}   
\centering
\includegraphics[width=\textwidth]{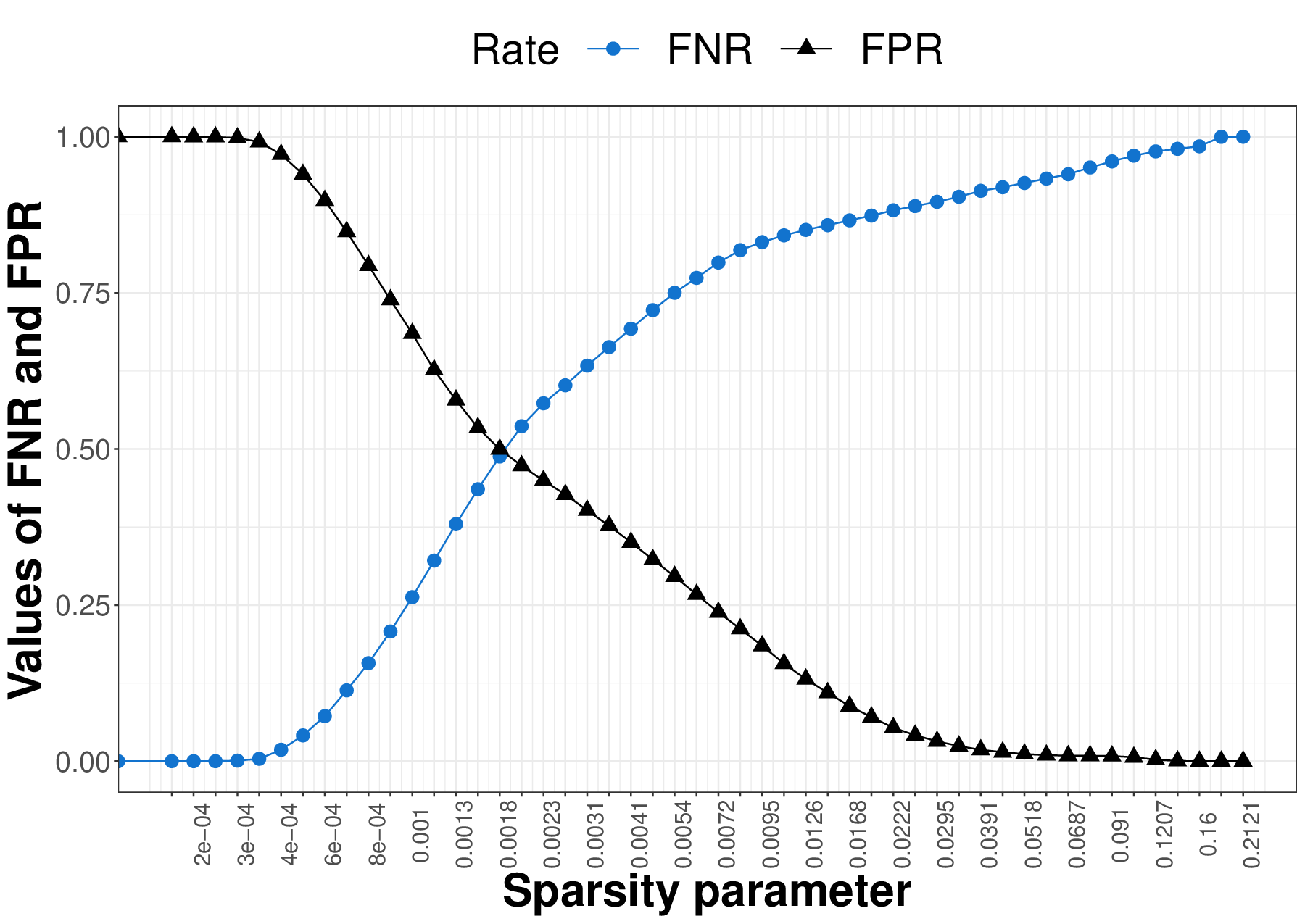}
\caption{Scenario C}
\label{3subfig:2f}
\end{subfigure}%
\caption{Simulation scenarios A, B and C for $D=20$. Left: percentage of zero logratios (black) and explained variability (red) for different values of the sparsity parameter $\alpha$. Right: FPR (black) and FNR (blue) for different values of $\alpha$.}
\label{fig:Sim2}
\end{figure}

Next, we zoom into the results of sparse PCA in Fig. \ref{fig:Sim2}. In scenario A, a lower percentage of explained variability is obtained for the dense model (compared to the $D=10$-part case), but it again decreases slightly with increased sparsity. The sparsity level rises even more rapidly than in the $D=10$-part case, achieving around 53\% of zero logratios ($\alpha = 0.001$) with roughly 42\% of explained variability. 
Similarly in scenario B, the sparsity level shoots up  quickly when the sparsity parameter increases. 
Finally for scenario C, we observe a similar trend (as for the $D=10$-case) in the decrease of explained variability with rising sparsity.

\end{document}